\documentclass[structabstract]{aa}  
\usepackage{graphicx}
\usepackage{txfonts}
\usepackage{natbib}
\usepackage{rotating}
\usepackage{longtable,lscape}

\begin{document}
   \title{Ba \& Eu Abundances in M15 giant stars\thanks{Based on observations collected at the European Organisation for Astronomical Research in the Southern Hemisphere, Chile. Proposal ID: 080.B-0784(A)}} 


   \author{C.~C.~Worley
          \inst{1,2}
          \and
	  V.~Hill
	  \inst{1}
          \and
          J.~Sobeck
	  \inst{1,3}
          \and
	  E.~Carretta
	  \inst{4}
          }

   \institute{1. Laboratoire Lagrange (UMR7293), Universit\'e de Nice Sophia Antipolis, CNRS, Observatoire de la C\^ote d'Azur, BP 4229, F-06304 Nice Cedex 04, France\\
              \email{ccworley@ast.cam.ac.uk} \\
	      2. Institute of Astronomy, University of Cambridge, Madingley Rd, Cambridge, CB3 0HA, U.K. \\
	      3. JINA: Joint Institute for Nuclear Astrophysics and the Department of Astronomy and Astrophysics, University of Chicago, 5640 South Ellis Avenue, Chicago, IL 60637, USA \\
              4. INAF – Osservatorio Astronomico di Bologna, via Ranzani 1, 40127, Bologna, Italy }

   \date{Received Jan, 2013; accepted Feb, 2013}

 
  \abstract
   {}
   {To investigate the Ba and Eu abundances for a sample of 63 giant stars in the globular cluster M15. This is the largest sample of M15 giants stars for which Ba abundances have been determined and, due to the target selection of the original research programme, the Ba abundances are complete along the red giant branch.}
   {Stellar parameters were taken from the previous key study and a microturbulence-surface gravity relation was determined for precise measurement of the Ba line at 6496.898~\AA, which has a high sensitivity to microturbulence. Element abundances for Ba, La, Eu, Ca, Ni and Fe were calculated using spectrum synthesis and equivalent widths techniques.}
   {A bimodal distribution in Ba, Eu and La abundances was found within the sample. The low Ba,Eu,La mode had mean abundances of $<$[Ba/H]$>$=--2.41$\pm$0.16, $<$[Eu/H]$>$=--1.80$\pm$0.08 and $<$[La/H]$>$=--2.19$\pm$0.13 while the high Ba,Eu,La mode had mean abundances of $<$[Ba/H]$>$=--2.00$\pm$0.16, $<$[Eu/H]$>$=--1.65$\pm$0.13 and $<$[La/H]$>$=--1.95$\pm$0.11.}
   {Both modes are indicative of a pollution scenario dominated by the {\it r}-process, hence contributions from explosive nucleosynthesis of massive stars. There may be evidence of further enhancement by another heavy element process and of potential anticorrelations in Na-O for both modes indicating a complex formation and evolution history for M15.}


   \maketitle
%

\section{Introduction}
The globular clusters of the Milky Way provide a wealth of information regarding the evolution of stars and the galaxy. Recent advances in technology and analysis techniques have revealed that many globular clusters show evidence of multiple stellar populations through photometry and chemical abundance distributions. A recent review of these discoveries is given in \citet{Gratton2012}. Disentangling the different stellar populations by their chemical signatures allows astronomers to identify the pollution and evolution mechanisms at work in these complex stellar systems.

The origins of many of the light (Z~$\leq$~30) and heavy (Z~$>$~30) elements can be attributed to key stages in stellar evolution, whereby each stage produces unique chemical signatures, depending on the mass and metallicity of the progenitor star. The contributions from the different stages of stellar evolution are determined by identifying the quantity and distribution of these elements in the observed stellar population and considering then in combination with the knowledge of the sites of nucleosynthesis in which the elements are produced.

The metal-poor globular cluster M15, with [Fe/H]=--2.31$\pm$0.06~dex \citep{Carretta2009a}, has proven to be an interesting case for study due in particular to star-to-star variations in [Eu/H] and [Ba/H] that have been found. Eu is predominantly an {\it r}-process element, formed due to the rapid capture of neutrons by a seed nuclei (Fe) in a high density neutron environment. The key stellar evolution nucleosynthesis site for the {\it r}-process is explosive nucleosynthesis in massive stars, such as in Type~II Supernov\ae\ \citep{Sneden2008}, which is also the key production site for the $\alpha$~elements (O, Mg, Si, Ca, Ti ...) \citep{Edvardsson1993}. Hence comparing Eu and $\alpha$~elements may be used to probe the mass-range of the supernov\ae\ progenitors. Ba is a neutron capture element that can be produced by the {\it r}-process, but is predominantly produced in the solar system material by the {\it s}-process, which is the slow capture of neutrons by a seed nuclei. The typical nucleosynthesis site 
for the {\it s}-process is within thermally pulsing asymptotic giant branch (AGB) stars \citep{Busso2001} and it is possibly also significant in massive fast-rotating stars \citep{Iliadis07book}. Hence the comparison of {\it s}- and {\it r}-process element abundances also provides a means by which to discriminate between the contributions from these very different nucleosynthesis sites.

The study of \citet[][hereafter, S97]{Sneden1997} investigated the abundances of both light (O, Na, Mg, Al, ...) and heavy (Ba, Eu) elements in a sample of 18 bright giants in M15. Anti-correlations typical to globular cluster were observed between Na and O, and between Al and Mg. Fe, Ca and Si were found to have uniform abundances for the sample. Eu and Ba were found to be strongly correlated with a possible bimodal distribution of both abundances, and to each have a large spread in values on the order of 0.6~dex. But no obvious correlation was found between these heavy elements and the light elements. Hence it was concluded that Ba and Eu had a primordial origin due to Type II Supernova of a stellar mass high enough to produce {\it r}-process elements but not sufficiently high to produce $\alpha$ elements, hence the lack of any correlation between those two types of elements.

\begin{figure*}[!th]
\begin{minipage}{180mm}  
\centering
\includegraphics[width=190mm]{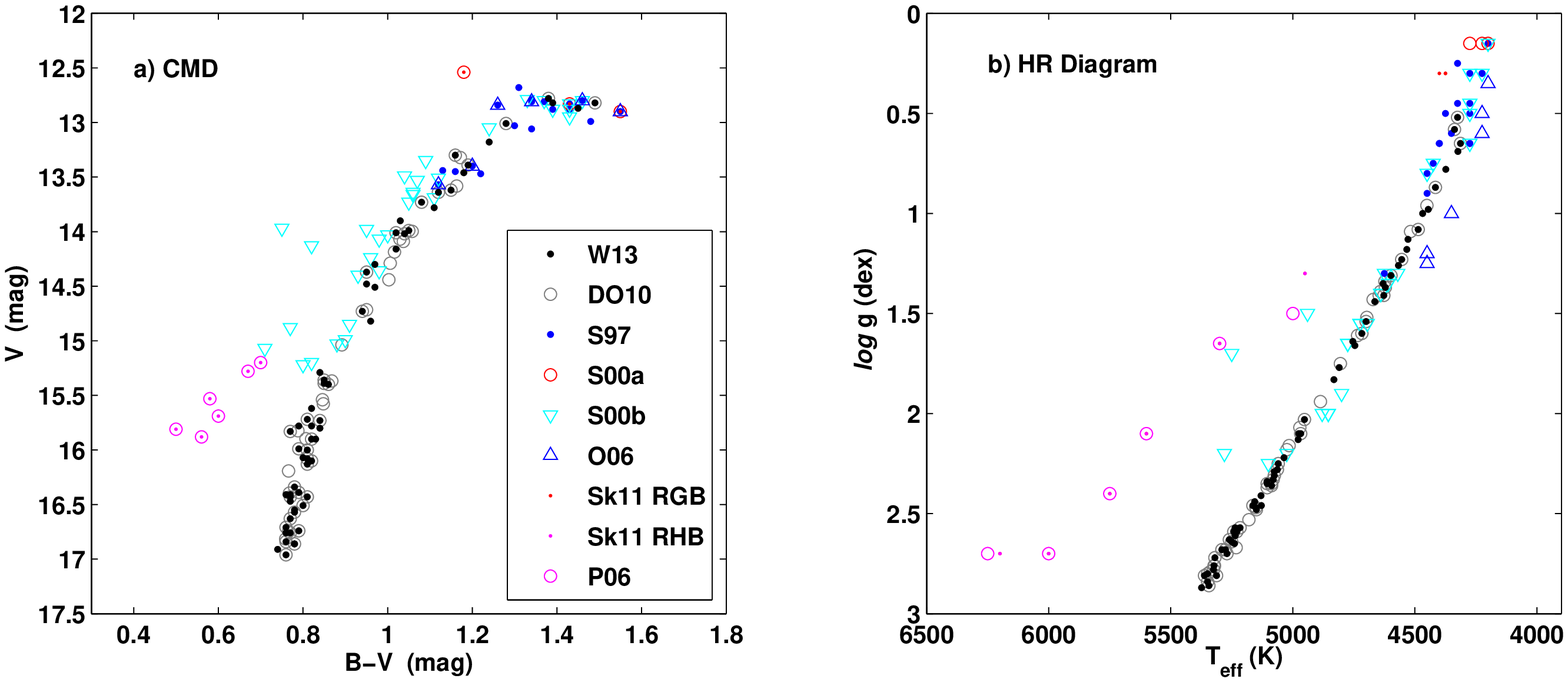}
\caption{The CMD and HR diagram of the key samples of giant stars in M15 as per the legend. The colours and parameters for W13 and DO10 both come from C09.}\label{fig:hrallpap}
\end{minipage}
\end{figure*}

This work was then followed by two further studies, \citet[][hereafter, S00a]{Sneden2000a} and \citet[][hereafter, S00b]{Sneden2000b}. In the former study three of the stars at the tip of the giant branch analysed in \citet{Sneden1997} were re-observed to investigate further the neutron-capture abundances in these stars. The stars were chosen as having similar stellar parameters but very different Eu and Ba abundances as per \citet{Sneden1997}. These differences were confirmed and the heavy element signatures that were derived closely followed that produced primarily by the {\it r}-process.

In the latter study a further 31 giants stars in M15 were analysed to investigate the proton-capture (Na), $\alpha$-capture (Ca, Si, Ti) and neutron-capture (Ba) element abundances. It was found that there was a large range in Na abundances, which was attributed to proton-capture on Ne occurring internally to the observed stars, as well as a large range in Ba comparable to that found in \citet{Sneden1997} confirming that result and the bimodal distribution of Ba. Eu abundances could not be measured for this sample.

\citet[][hereafter, O06]{Otsuki2006} investigated the heavy element abundances of seven giant stars in M15 that had also been analysed in S97. They found Fe and Eu abundances in agreement with S97 and extended the analysis to Zr, Y and La. They found that La correlated with Eu, as expected, but observed an anti-correlation of Zr and Y with Eu. From this they concluded that while the La (and Ba) correlation with Eu implies a primary contribution of neutron-capture elements by an r-process only mechanism (e.g. explosive nucleosynthesis within massive stars) this cannot also explain the anticorrelation of Zr and Y with Eu. They argue that some other process enriched the inter-stellar medium in M15 in light neutron-capture elements before the enrichment of the heavy neutron-capture elements by supernova.

\citet[][hereafter, P06]{Preston2006} made an investigation of six red horizontal branch (RHB) stars in M15. Both light and heavy element abundances were determined, for the M15 stars similar results regarding the evidence of star-to-star scatter in the heavy element abundances were found. Also noted was that the RHB stars of M15 were found to be $\sim$0.2~dex more metal-poor than red giant branch (RGB) stars in the same cluster.

Most recently, a detailed investigation into neutron-capture element abundances in giant stars in M15 was carried out by \citet[][hereafter, Sk11]{Sobeck2011}. The three RGB stars of S00a and the six RHB stars of P06 were re-analysed in a consistent manner in order to investigate the nature of the heavy element abundances in M15 giants stars, and to investigate the metallicity offset between the RHB and RGB as observed in P06. The abundances for 40 species of both light and heavy elements were derived for these nine stars. As had been observed previously the abundances of proton-capture elements showed no correlation with the abundances of the neutron capture elements. In particular a metallicity offset between the RHB and RGB was found, and the large spread in Ba and Eu was confirmed. However the bimodal distribution of Ba and Eu that was proposed in \citet{Sneden1997} was not observed. The analysis concluded that the heavy neutron-capture element abundances agreed well with a scaled-solar {\it r}-process 
distribution, but the light neutron-capture element abundances did not agree with either a scaled-solar {\it s}- or {\it r}-process distribution. This was therefore evidence of some other nucleosynthetic process, under the umbrella of the Lighter Element Primary Process (LEPP), at work.

The largest sample of M15 stars analysed spectroscopically to-date was carried out by \citet[][hereafter, C09]{Carretta2009a} as part of an analysis of light element abundances for stellar samples within 15 galactic globular clusters observed using FLAMES/GIRAFFE. Fe, Na and O abundances were investigated for 84 stars in M15. In total the research programme obtained homogeneously determined abundances for Na and O for 1235 stars in 19 globular clusters. An anti-correlation between Na and O was observed for all the clusters, including M15. This provided evidence of a pristine first generation of stars ($\sim$30\% of the observed cluster stars) and a second generation that had formed from material enriched in H-burning products by now-extinct massive stars of the first stellar generation. \citet{Carretta2009b} followed up this study with high resolution observations of 200 red giants across 17 GCs, for which 13 stars were observed in M15 from the C09 sample.

\begin{table*}[!htp]
\caption{The star ID, effective temperature, surface gravity, model metallicity, microturbulence, radial velocity, abundances derived from Fe~I and Fe~II lines, Na and O abundances and their associated root mean square error (rms) for each of the M15 stars analysed here as determined in C09}
\begin{center}
\begin{tabular}{cccccccccccccc}
\hline\hline
ID & $T_{\textrm{eff}}$ & $\log g$ & [A/H] & $\xi$ & V$_{rad}$ & [Fe~I/H] & rms & [Fe~II/H] & rms & [O/Fe] & rms & [Na/Fe] & rms \\ 
 & (K) & (dex) & (dex) & (kms$^{-1}$) & (kms$^{-1}$) & (dex) &  & (dex) &  & (dex) &  & (dex) &  \\ 
\hline
40825 & 4313 & 0.65 & --2.33 & 2.25 & --113.91 & --2.33 & 0.14 & --2.34 & - & 0.24 & 0.11 & 0.44 & 0.07 \\ 
4099 & 4324 & 0.69 & --2.32 & 2.03 & --109.43 & --2.32 & 0.11 &  &  & 0.54* & 0.13 & --0.07* & 0.01 \\ 
43788 & 4325 & 0.52 & --2.36 & 2.40 & --96.93 & --2.36 & 0.10 & --2.46 & 0.08 & 0.39 & 0.02 & 0.13 & 0.06 \\ 
31914 & 4338 & 0.58 & --2.20 & 2.09 & --112.77 & --2.20 & 0.10 & --2.21 & - & -0.32 & - &  &  \\ 
41287 & 4373 & 0.78 & --2.36 & 1.10 & --116.25 & --2.36 & 0.07 &  &  &  &  & 0.39 & 0.01 \\ 
34519 & 4416 & 0.87 & --2.38 & 1.62 & --98.04 & --2.38 & 0.11 & --2.46 & - & 0.46 & 0.04 & 0.11 & 0.08 \\ 
37215 & 4445 & 0.98 & --2.31 & 1.28 & --111.36 & --2.31 & 0.01 &  &  &  &  & 0.73 & - \\ 
34995 & 4468 & 1.00 & --2.34 & 0.79 & --115.14 & --2.34 & 0.08 &  &  &  &  & 0.64 & - \\ 
3137 & 4486 & 1.08 & --2.35 & 2.27 & --109.50 & --2.35 & 0.13 & --2.29 & - & 0.29 & 0.08 & 0.58 & 0.04 \\ 
42262 & 4528 & 1.13 & --2.33 & 2.50 & --104.50 & --2.33 & 0.08 &  &  &  &  & 0.90 & 0.13 \\ 
26751 & 4533 & 1.18 & --2.44 & 1.04 & --98.72 & --2.44 & 0.08 &  &  &  &  & 0.55 & 0.06 \\ 
38678 & 4554 & 1.23 & --2.35 & 1.28 & --109.01 & --2.35 & 0.08 & --2.46 & 0.01 & 0.13 & - & 0.56 & 0.03 \\ 
2792 & 4567 & 1.26 & --2.32 & 1.50 & --100.93 & --2.32 & 0.13 &  &  & 0.32* & 0.00 & 0.20* & 0.06 \\ 
39752 & 4598 & 1.31 & --2.36 & 1.30 & --108.38 & --2.36 & 0.13 & --2.39 & 0.03 & 0.26 & - & 0.29 & 0.15 \\ 
34456 & 4621 & 1.37 & --2.34 & 1.40 & --97.02 & --2.34 & 0.20 &  &  & 0.22 & - & 0.34 & 0.01 \\ 
34961 & 4627 & 1.41 & --2.32 & 1.85 & --113.28 & --2.32 & 0.15 & --2.36 & 0.02 & -0.01 & - & 0.65 & 0.03 \\ 
39787 & 4630 & 1.35 & --2.43 & 2.10 & --111.94 & --2.43 & 0.13 &  &  & 0.48* & 0.00 & 0.13* & 0.03 \\ 
38329 & 4664 & 1.44 & --2.37 & 0.86 & --109.04 & --2.37 & 0.02 &  &  &  &  & 0.16 & - \\ 
45062 & 4700 & 1.54 & --2.31 & 1.66 & --110.27 & --2.31 & 0.11 & --2.23 & - & 0.03 & - & 0.63 & 0.05 \\ 
20498 & 4717 & 1.61 & --2.49 & 1.12 & --106.58 & --2.49 & 0.08 & --2.44 & - & 0.25 & - & 0.54 & 0.04 \\ 
21948 & 4746 & 1.66 & --2.49 & 0.20 & --108.68 & --2.49 & 0.05 &  &  &  &  & 0.46 & - \\ 
28510 & 4754 & 1.64 & --2.42 & 0.14 & --100.52 & --2.42 & 0.02 &  &  & 0.23* & 0.00 & 0.07* & 0.07 \\ 
29264 & 4810 & 1.77 & --2.35 & 0.84 & --101.86 & --2.35 & 0.14 &  &  & 0.40* & 0.00 & 0.01* & 0.00 \\ 
18815 & 4832 & 1.83 & --2.29 & 0.80 & --108.68 & --2.29 & 0.10 &  &  & 0.66* & 0.00 & 0.00* & 0.00 \\ 
42362 & 4952 & 2.03 & --2.41 & 1.00 & --110.67 & --2.41 & 0.09 &  &  & 0.27 & - & 0.38 & 0.17 \\ 
29480 & 4968 & 2.10 & --2.33 & 1.16 & --104.60 & --2.33 & 0.16 &  &  & 0.30 & - & 0.45 & 0.14 \\ 
37931 & 4976 & 2.10 & --2.37 & 1.21 & --111.25 & --2.37 & 0.05 &  &  &  &  & 0.27 & 0.08 \\ 
31791 & 4978 & 2.13 & --2.28 & 0.10 & --104.92 & --2.28 & 0.03 &  &  &  &  & 0.14 & - \\ 
33484 & 5036 & 2.22 & --2.44 & 1.20 & --109.75 & --2.44 & 0.17 &  &  &  &  & 0.14 & - \\ 
10329 & 5060 & 2.25 & --2.38 & 1.25 & --109.91 & --2.38 & 0.11 &  &  & 0.42 & - & 0.53 & - \\ 
27889 & 5063 & 2.28 & --2.32 & 1.15 & --101.14 & --2.32 & 0.13 &  &  & 0.63 & - & 0.04 & 0.11 \\ 
39741 & 5076 & 2.31 & --2.25 & 1.42 & --112.10 & --2.25 & 0.20 &  &  & 0.43 & - & 0.38 & 0.18 \\ 
42674 & 5076 & 2.29 & --2.43 & 1.35 & --96.28 & --2.43 & 0.06 &  &  &  &  & 0.67 & - \\ 
17458 & 5081 & 2.33 & --2.31 & 0.92 & --110.56 & --2.31 & 0.19 &  &  & 0.44 & - & 0.16 & 0.05 \\ 
40086 & 5087 & 2.36 & --2.32 & 1.48 & --122.86 & --2.32 & 0.10 &  &  & 0.41 & - & 0.37 & 0.07 \\ 
36274 & 5105 & 2.35 & --2.35 & 1.16 & --100.30 & --2.35 & 0.13 &  &  & 0.50 & - & 0.34 & 0.04 \\ 
26759 & 5106 & 2.34 & --2.35 & 1.16 & --101.07 & --2.35 & 0.20 &  &  &  &  & 0.24 & - \\ 
29521 & 5129 & 2.46 & --2.30 & 0.03 & --101.45 & --2.30 & 0.02 &  &  &  &  & 0.19 & 0.08 \\ 
41987 & 5130 & 2.41 & --2.33 & 2.20 & --108.07 & --2.33 & 0.09 &  &  &  &  & 0.39 & - \\ 
1939 & 5149 & 2.48 & --2.41 & 0.88 & --101.98 & --2.41 & 0.03 &  &  & 0.32 & - &  &  \\ 
34272 & 5151 & 2.47 & --2.25 & 1.24 & --108.35 & --2.25 & 0.10 &  &  & 0.47 & - & 0.38 & 0.13 \\ 
35961 & 5156 & 2.44 & --2.41 & 0.40 & --107.90 & --2.41 & 0.10 &  &  &  &  & 0.38 & 0.06 \\ 
31010 & 5164 & 2.46 & --2.32 & 0.66 & --115.45 & --2.32 & 0.08 &  &  &  &  & 0.61 & 0.03 \\ 
23216 & 5216 & 2.57 & --2.32 & 0.14 & --104.01 & --2.32 & 0.02 &  &  &  &  &  &  \\ 
18240 & 5229 & 2.59 & --2.30 & 0.94 & --103.56 & --2.30 & 0.03 &  &  &  &  & 0.56 & 0.20 \\ 
32942 & 5236 & 2.61 & --2.25 & 1.82 & --108.90 & --2.25 & 0.10 &  &  &  &  & 0.38 & 0.04 \\ 
32979 & 5236 & 2.57 & --2.35 & 1.97 & --106.56 & --2.35 & 0.04 &  &  &  &  & 0.17 & - \\ 
2411 & 5239 & 2.65 & --2.29 & 1.07 & --110.01 & --2.29 & 0.22 &  &  &  &  &  &  \\ 
23153 & 5242 & 2.59 & --2.26 & 1.61 & --108.68 & --2.26 & 0.10 &  &  &  &  &  &  \\ 
8927 & 5250 & 2.64 & --2.42 & 0.16 & --104.84 & --2.42 & 0.08 &  &  &  &  & 0.30 & 0.10 \\ 
18770 & 5260 & 2.63 & --2.38 & 1.95 & --103.56 & --2.38 & 0.10 &  &  &  &  &  &  \\ 
22441 & 5270 & 2.70 & --2.35 & 0.35 & --108.68 & --2.35 & 0.09 &  &  &  &  &  &  \\ 
18508 & 5276 & 2.68 & --2.26 & 2.61 & --103.56 & --2.26 & 0.09 &  &  &  &  &  &  \\ 
9608 & 5291 & 2.68 & --2.33 & 0.31 & --104.47 & --2.33 & 0.12 &  &  & 0.41 & - &  &  \\ 
29436 & 5312 & 2.81 & --2.41 & 1.39 & --98.38 & --2.41 & 0.11 &  &  &  &  &  &  \\ 
4989 & 5319 & 2.72 & --2.38 & 0.10 & --104.70 & --2.38 & 0.06 &  &  &  &  &  &  \\ 
3825 & 5323 & 2.76 & --2.26 & 0.72 & --109.12 & --2.26 & 0.12 & --2.24 & 0.34 &  &  &  &  \\ 
40762 & 5325 & 2.78 & --2.34 & 1.17 & --115.00 & --2.34 & 0.07 & --2.21 & - &  &  & 0.66 & - \\ 
8329 & 5344 & 2.86 & --2.41 & 1.77 & --106.98 & --2.41 & 0.02 &  &  &  &  &  &  \\ 
2463 & 5349 & 2.84 & --2.31 & 1.01 & --105.89 & --2.31 & 0.12 & --2.27 & - & 0.48 & - & 0.40 & 0.08 \\ 
19346 & 5349 & 2.80 & --2.34 & 2.70 & --105.99 & --2.34 & 0.04 &  &  &  &  & 0.94 & 0.03 \\ 
28026 & 5362 & 2.81 & --2.38 & 1.80 & --111.48 & --2.38 & 0.15 &  &  &  &  &  &  \\ 
31796 & 5374 & 2.87 & --2.25 & 1.80 & --110.52 & --2.25 & 0.08 &  &  &  &  & 0.08 & - \\ 
 &  &  &  &  &  &  &  &  &  &  &  &  &  \\ 
 & \multicolumn{ 8}{l}{* [O/Fe] \& [Na/Fe] taken from \citet{Carretta2009b}} &  & & & &  \\
\hline
\end{tabular}
\end{center}
\label{tab:m15_c09}
\end{table*}

Of particular relevance is the study undertaken in \citet[][hereafter, DO10]{DOrazi2010} which investigated Ba abundances in GC stars directly from the observations of C09. This study measured Ba for 1200 stars across the 15 GCs, including measurements for 57 stars in the M15 sample. M15 was the only sample for which they found a large variation in the Ba abundances between the stars confirming the large variation found in S97 and S00b.

The study carried out here seeks to extend the sample of giant stars in M15 for which heavy neutron-capture element abundances have been determined and consider their distribution in M15 in light of the previous studies.

Section~\ref{sec:obs} describes the observations and presents the stellar parameters for the dataset under analysis. Section~\ref{sec:stellarparams} describes the determination of the stellar parameters as per \citet{Carretta2009a} and the treatment of the microtubulence, $\xi$, in this study. Section~\ref{sec:stellarabund} describes the determination of the chemical abundances for this sample. Section~\ref{sec:stellarabundanalysis} present the results of the chemical abundance analysis. Section~\ref{sec:heavylightdis} discusses possible scenarios based on the results. Section~\ref{sec:conclusion} concludes the paper.

\section{Observations}\label{sec:obs}
Observations of 63 red giant branch (RGB) stars in M15 were taken using the HR14A setup of GIRAFFE on the VLT in November 2009. The spectra were wavelength calibrated as from the ESO pipeline and sky-subtraction was performed using IRAF. Normalisation and radial velocity determination were performed using tools developed for the AMBRE Project \citep{Worley2012}. The signal-to-noise (S/N) ranged from 50 to 170 across the sample.

These targets were taken from C09 in which the stellar parameters of effective temperature ($T_{\textrm{eff}}$), surface gravity ($\log g$) and metallicity ([Fe/H]) as well as individual chemical abundances for Na and O were obtained for a total sample of 84 stars in M15. Table~\ref{tab:m15_c09} lists the star ID and stellar parameters of the 63 stars from C09 that were analysed here as taken from Table~3 in C09. 

The DO10 sample was also a subsample of C09. Of the 57 DO10 stars, 38 were also observed for this study. D010 could not analyse the complete C09 sample for Ba abundances as not all the stars were observed using the HR13 grating, in which the Ba spectral feature at 6141.73~\AA\ is present. 

The colour-magnitude diagram (CMD) and Hertsprung-Russell (HR) diagram of the 63 stars from C09 that have been observed and analysed here are shown in Figure~\ref{fig:hrallpap}. The locations of stars from previous key studies of M15 are also shown for comparison including the DO10 subsample of C09. 

The wavelength range of the HR14 grating covers from $\sim$6308~\AA\ to $\sim$6700~\AA, which includes the key neutron-capture features of Ba~II at 6496.898~\AA, La~II at 6390.4770~\AA, and Eu~II at 6437.640~\AA\ and 6645.127~\AA. At this resolution of R$\sim$14,500 these three features are observable although the Eu~II features are weak. The astrophysical quantities associated with these lines, as well as those for the other lines measured in this study, are listed in Table~\ref{tab:linelist}. Table~\ref{tab:hfs} lists the hyperfine structure decomposition for the Ba~II, La~II and Eu~II spectral lines as well as the references from which the atomic information was obtained upon which the calculations were based.

\begin{table}[htbp]
\caption{List of Fe, Ca, Ni, Ba, Eu and La lines used in this analysis with the associated excitation potential ($\chi$), $\log gf$ and method of analysis used either spectrum synthesis (SS) or equivalent width curve of growth (EW). The reference solar abundances are also listed.}
\begin{tabular}{ccccccc}
\hline\hline
{\scriptsize Wavelength (\AA)} & {\scriptsize Species} & $\chi$ (eV) & {\scriptsize $\log gf$} & {\scriptsize Reference} & {\scriptsize Method} & {\scriptsize Sol*} \\ 
\hline
6439.0700 & 20.0 & 2.526 & 0.470 & NIST & SS & 6.41 \\ 
6471.6600 & 20.0 & 2.526 & --0.590 & NIST & SS & `` \\ 
6318.0171 & 26.0 & 2.453 & --1.804 & NIST & EW & 7.54 \\ 
6322.6851 & 26.0 & 2.588 & --2.426 & NIST & EW & `` \\ 
6335.3304 & 26.0 & 2.198 & --2.177 & NIST & EW & `` \\ 
6336.8239 & 26.0 & 3.686 & --0.856 & NIST & EW & `` \\ 
6344.1487 & 26.0 & 2.433 & --2.923 & NIST & EW & `` \\ 
6355.0286 & 26.0 & 2.845 & --2.291 & NIST & EW & `` \\ 
6393.6009 & 26.0 & 2.433 & --1.576 & NIST & EW & `` \\ 
6408.0179 & 26.0 & 3.686 & --1.018 & NIST & EW & `` \\ 
6411.6489 & 26.0 & 3.654 & --0.718 & NIST & EW & `` \\ 
6421.3504 & 26.0 & 2.279 & --2.027 & NIST & EW & `` \\ 
6430.8460 & 26.0 & 2.176 & --2.006 & NIST & EW & `` \\ 
6475.6240 & 26.0 & 2.559 & --2.942 & NIST & EW & `` \\ 
6481.8699 & 26.0 & 2.279 & --2.984 & NIST & EW & `` \\ 
6518.3667 & 26.0 & 2.832 & --2.298 & NIST & EW & `` \\ 
6575.0154 & 26.0 & 2.588 & --2.710 & NIST & EW & `` \\ 
6592.9134 & 26.0 & 2.728 & --1.473 & NIST & EW & `` \\ 
6593.8701 & 26.0 & 2.433 & --2.422 & NIST & EW & `` \\ 
6432.6800 & 26.1 & 2.891 & --3.500 & NIST & EW & `` \\ 
6456.3760 & 26.1 & 3.903 & --2.190 & NIST & EW & `` \\ 
6516.0530 & 26.1 & 2.891 & --3.372 & NIST & EW & `` \\ 
6643.6300 & 28.0 & 1.675 & --2.300 & NIST & SS & 6.29 \\ 
6496.8980 & Ba II & 0.604 & --0.407 & {\small D12~$^{a}$} & SS & 2.25  \\ 
6390.4770 & La II & 0.321 & --1.410 & {\small L01a~$^{b}$} & SS & 1.25  \\ 
6437.6400 & Eu II & 1.320 & --0.320 & {\small L01b~$^{c}$} & SS & 0.60  \\ 
6645.1270 & Eu II & 1.380 & 0.120 & {\small L01b~$^{c}$} & SS & `` \\
\\ 
\multicolumn{7}{l}{{\scriptsize * Solar reference abundances taken from \citet{Lodders2003}}} \\  
\multicolumn{7}{l}{{\scriptsize a) \citet{Davidson1992}; b) \citet{Lawler2001a}; c) \citet{Lawler2001b}}}\\
\hline
\end{tabular}
\label{tab:linelist}
\end{table}

\begin{table*}[!htbp]
\centering
\caption{Hyperfine structure decomposition based on atomic information obtained from the associated references for the Ba II \citep{Rutten1978,Davidson1992}, La II \citep{Lawler2001a} and Eu II \citep{Lawler2001b} spectral lines used in the spectrum synthesis of the M15 giant stars.}
\begin{tabular}{ccc|cc}
\hline\hline
Wavelength($\AA$) & Strength & & Wavelength($\AA$) & Strength \\ 
\hline
\multicolumn{2}{l}{La II $\lambda$6496.898: $\chi$=0.321~eV, $\log gf$=--1.410} & &  \multicolumn{2}{l}{Ba~II $\lambda$6496.898: $\chi$=0.604~eV, $\log gf$=--0.407} \\
\hline
\multicolumn{1}{l}{$^{139}$La:} & & & \multicolumn{1}{l}{$^{134}$Ba:} &  \\ 
6390.455............. & --2.012 & & 6496.898............. & --0.406 \\ 
6390.468............. & --2.183 & & \multicolumn{1}{l}{$^{135}$Ba:} &  \\ 
6390.468............. & --2.752 & & 6496.887............. & --1.911 \\ 
6390.479............. & --2.570 & & 6496.890............. & --1.212 \\ 
6390.479............. & --3.752 & & 6496.895............. & --0.765 \\ 
6390.480............. & --2.390 & & 6496.903............. & --1.610 \\ 
6390.489............. & --2.536 & & 6496.905............. & --1.212 \\ 
6390.489............. & --3.334 & & 6496.908............. & --1.212 \\ 
6390.490............. & --2.661 & & \multicolumn{1}{l}{$^{136}$Ba:} &  \\ 
6390.496............. & --3.100 & & 6496.898............. & --0.406 \\ 
6390.497............. & --2.595 & & \multicolumn{1}{l}{$^{137}$Ba:} &  \\ 
6390.498............. & --3.079 & & 6496.886............. & --1.911 \\ 
6390.502............. & --2.954 & & 6496.889............. & --1.212 \\ 
6390.503............. & --2.778 & & 6496.894............. & --0.765 \\ 
6390.506............. & --2.857 & & 6496.904............. & --1.610 \\ 
                      &         & & 6496.906............. & --1.212 \\ 
                      &         & & 6496.909............. & --1.212 \\
                      &         & & \multicolumn{1}{l}{$^{138}$Ba:} &  \\
                      &         & & 6496.898............. & --0.406 \\
\hline
\multicolumn{2}{l}{Eu~II $\lambda$6437.640: $\chi$=1.320~eV, $\log gf$=--0.320}  & & \multicolumn{2}{l}{Eu II $\lambda$6645.127: $\chi$=1.380~eV, $\log gf$=+0.120} \\
\hline
\multicolumn{1}{l}{$^{151}$Eu:} &  & & \multicolumn{1}{l}{$^{151}$Eu:} & \\ 
6437.611............. & --0.960 &  & 6645.071............. & --0.517   \\ 
6437.619............. & --2.191 &  & 6645.078............. & --1.823   \\ 
6437.629............. & --2.191 &  & 6645.085............. & --3.467 \\ 
6437.636............. & --1.070 &  & 6645.097............. & --0.593 \\ 
6437.643............. & --1.998 &  & 6645.104............. & --1.628 \\ 
6437.650............. & --1.998 &  & 6645.112............. & --3.149 \\ 
6437.656............. & --1.181 &  & 6645.118............. & --0.672 \\ 
6437.662............. & --1.956 &  & 6645.126............. & --1.583 \\ 
6437.666............. & --1.956 &  & 6645.133............. & --3.077 \\ 
6437.672............. & --1.287 &  & 6645.137............. & --0.754 \\ 
6437.676............. & --2.010 &  & 6645.144............. & --1.635 \\ 
6437.679............. & --2.010 &  & 6645.150............. & --3.245 \\ 
6437.684............. & --1.377 &  & 6645.152............. & --0.839 \\ 
6437.687............. & --2.207 &  & 6645.158............. & --1.830 \\ 
6437.689............. & --2.207 &  & 6645.163............. & --0.921 \\ 
6437.692............. & --1.428 &  & \multicolumn{1}{l}{$^{153}$Eu:} & \\ 
\multicolumn{1}{l}{$^{153}$Eu:} & & & 6645.072............. & --1.823  \\ 
6437.610............. & --0.960 &  & 6645.074............. & --0.517\\ 
6437.613............. & --2.191 &  & 6645.074............. & --3.467\\ 
6437.624............. & --2.191 &  & 6645.087............. & --0.593 \\ 
6437.627............. & --1.070 &  & 6645.089............. & --1.628\\ 
6437.630............. & --1.998 &  & 6645.094............. & --3.149\\ 
6437.634............. & --1.998 &  & 6645.096............. & --0.672\\ 
6437.637............. & --1.181 &  & 6645.101............. & --1.583\\ 
6437.639............. & --1.956 &  & 6645.104............. & --0.754 \\ 
6437.640............. & --1.956 &  & 6645.107............. & --3.077 \\ 
6437.641............. & --2.010 &  & 6645.109............. & --0.839 \\ 
6437.642............. & --1.287 &  & 6645.110............. & --1.635 \\ 
6437.642............. & --2.207 &  & 6645.113............. & --0.921 \\ 
6437.643............. & --1.428 &  & 6645.116............. & --1.830 \\ 
6437.644............. & --1.377 &  & 6645.116............. & --3.245 \\ 
6437.644............. & --2.010 &  &  & \\ 
6437.645............. & --2.207 &  &  & \\ 
\hline
\end{tabular}
\label{tab:hfs}
\end{table*}

\section{Stellar Parameters}\label{sec:stellarparams}
The stellar parameters for each of the stars observed here had been determined in C09. The  $T_{\textrm{eff}}$ was determined in a two step process whereby a relation between $T_{\textrm{eff}}$ and the V (or K) magnitude was developed for the cluster using $V-K$ photometry and the \cite{Alonso1999} calibration. This relation was then used to calculate the $T_{\textrm{eff}}$ for each star in the sample. The $\log g$ was determined using $T_{\textrm{eff}}$, the apparent magnitude, the distance modulus and bolometric corrections from \cite{Alonso1999}. The $\xi$ was determined by eliminating any trend of the abundance of iron with expected line strength based on the equivalent width measurement of Fe~I lines. See C09 and references therein for more detailed descriptions of the methods employed.

However, as can be noted from Table~\ref{tab:m15_c09}, the resulting microturbulence velocities for the C09 M15 sample range from 0.03 to 2.7~kms$^{-1}$ which is a very wide and rather implausible range for this sample of giant stars. This wide range most probably results from too few and too weak Fe lines being available in the spectral range used in the C09 study for this very metal-poor cluster. These $\xi$ values anyway have almost no impact on the resulting abundances of Fe, O and Na investigated in C09, since the lines analysed were on the linear part of the curve of growth. On the other hand, the Ba line under analysis here is strong enough for the microturbulence to play a role in desaturating the lines.

The sensitivity of the 6496.898~\AA\ Ba line to $\xi$ is clearly shown in the error analysis in Table~\ref{tab:param_error}. In a preliminary analysis using the $\xi$ values determined in C09, a clear trend of Ba abundance with $\xi$ was found. This is a typical problem when dealing with the measurement of strong lines such as those of Ba in giants. In S00b a constant $\xi$ of 2.0~kms$^{-1}$ was assumed while Sk11 treated $\xi$ so as to eliminate any trend for the strong lines. 

For this sample we therefore re-determined $\xi$ using the available Fe~I lines as listed in Table~\ref{tab:linelist}. Hence the equivalent widths for Fe~I and Fe~II lines were measured for those stars in which a sufficient number of unblended lines could be identified. For the majority of stars the combination of low S/N and high temperatures meant a insufficient sample of Fe lines could be identified.

\begin{table*}[htbp]
\caption{Stellar Parameters from C09 for 13 stars for which the equivalent widths of Fe~I and Fe~II lines were measured and/or were analysed in previous studies as listed. The mean and standard deviation and the number of lines measured for Fe~I and Fe~II are listed. The standard deviation is given only for measurements of N$>$2.}
\begin{tabular}{cccccccccccccl}
\hline\hline
Star ID & Other ID & $T_{\textrm{eff}}$ & $\log g$ & [A/H] & $\xi$ & $\xi_{W13}$ & [Fe I/H] & $\sigma$ & N$^\#$ & [Fe II/H] & $\sigma$ & N & Other Studies \\ 
\hline
40825 & K386 & 4313 & 0.65 & --2.33 & 2.25 & 2.25 & --2.40 & 0.10 & 35 (22) & --2.37 & - & 2 & S97, S00b, Ot06 \\ 
4099 & K341 & 4324 & 0.69 & --2.32 & 2.03 & 2.30 & --2.40 & 0.11 & 17 & --2.34 & - & 2 & S97, S00a, S00b, Sk11 \\ 
43788 & K825 & 4325 & 0.52 & --2.36 & 2.40 & 2.25 & --2.32 & 0.12 & 17 & --2.37 & - & 2 & S97, S00b \\ 
31914 & K757 & 4338 & 0.58 & --2.20 & 2.90 & 2.10 & --2.27 & 0.11 & 17 & --2.33 & - & 2 & S97, S00b \\ 
41287 & K702 & 4373 & 0.78 & --2.36 & 1.10 & 2.10 & --2.41 & 0.13 & 17 & --2.37 & - & 2 & S97 \\ 
34519 &  & 4416 & 0.87 & --2.38 & 1.62 & 2.00 & --2.44 & 0.10 & 23 (9) & --2.32 & - & 2 &  \\ 
34995 &  & 4468 & 1.00 & --2.34 & 0.79 & 2.10 & --2.41 & 0.13 & 15 & --2.40 & - & 2 &  \\ 
3137 & K387 & 4486 & 1.08 & --2.35 & 2.27 & 2.00 & --2.38 & 0.12 & 16 & --2.32 & - & 2 & S97 \\ 
26751 & K709 & 4533 & 1.18 & --2.44 & 1.40 & 2.02 &  &  &  &  &  &  & S00b \\ 
38678 & K288 & 4554 & 1.23 & --2.35 & 1.28 & 2.00 & --2.47 & 0.11 & 20 (9) & --2.38 & - & 2 & S00b \\ 
2792 & K255 & 4567 & 1.26 & --2.32 & 1.50 & 2.20 & --2.47 & 0.13 & 11 & --2.37 & 0.14 & 3 & S00b \\ 
34961 &  & 4627 & 1.41 & --2.32 & 1.85 & 2.00 & --2.46 & 0.13 & 22 (14) & --2.46 & 0.09 & 3 &  \\ 
38329 &  & 4664 & 1.44 & --2.37 & 0.86 & 1.80 & --2.41 & 0.12 & 11 & --2.28 & - & 2 &  \\ 
 &  &  &  &  &  &  &  &  &  &  &  &  &  \\ 
\multicolumn{14}{l}{$\#$ The numbers in brackets are the number of equivalent widths obtained from C09}  \\ 
\hline
\end{tabular}
\label{tab:meaEW_feifeii}
\end{table*}

\begin{figure*}[!th]
\centering
\begin{minipage}{90mm}  
\includegraphics[width=95mm]{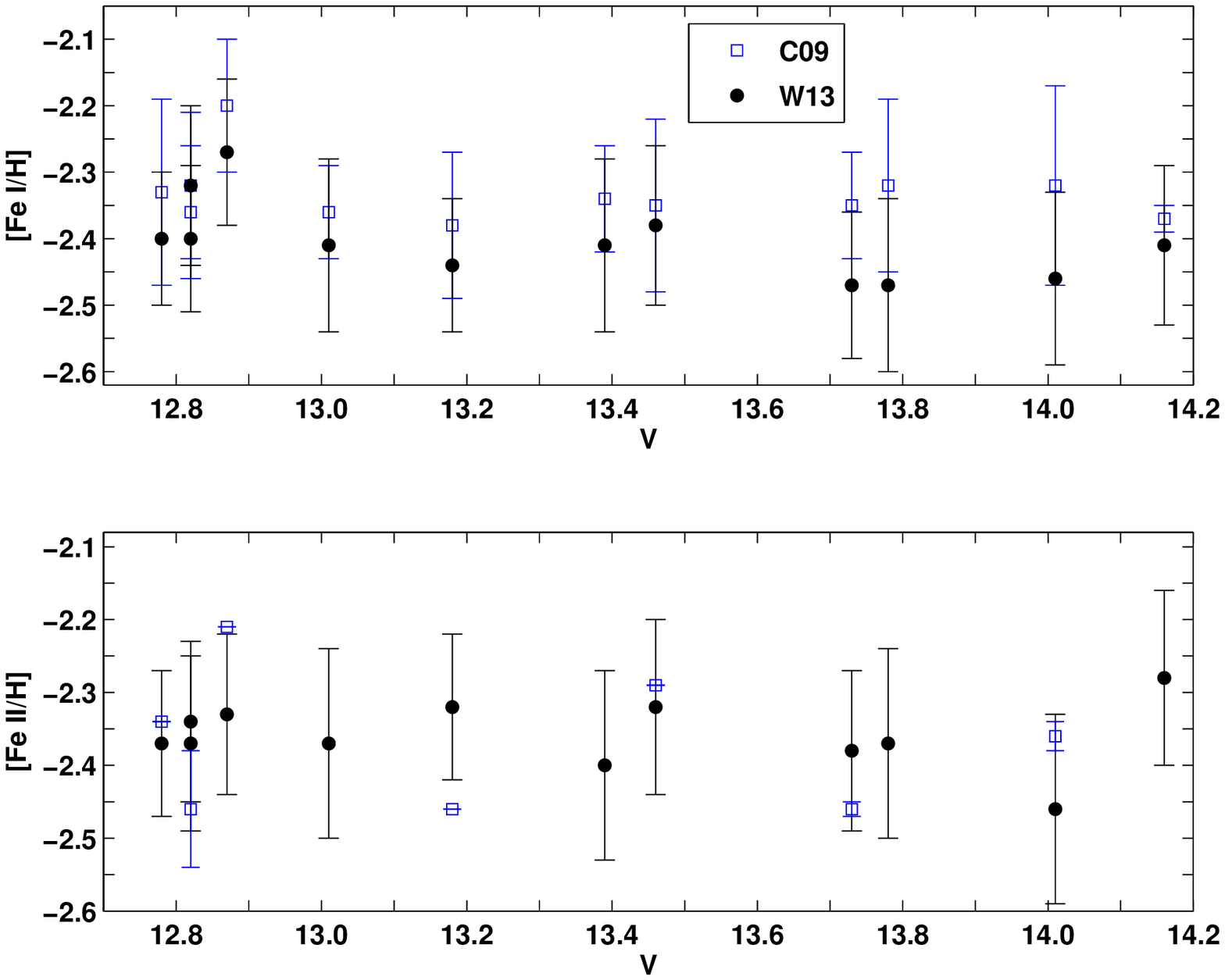}
\caption{Comparison of [Fe~I/H] and [Fe~II/H] values from C09 and the values determined here by equivalent width measurement. The C09 [Fe~I/H] are those used as input for the stellar atmosphere models.}\label{fig:fe1fe2comp}
\end{minipage}
\hfill
\begin{minipage}{90mm}  
\includegraphics[width=95mm]{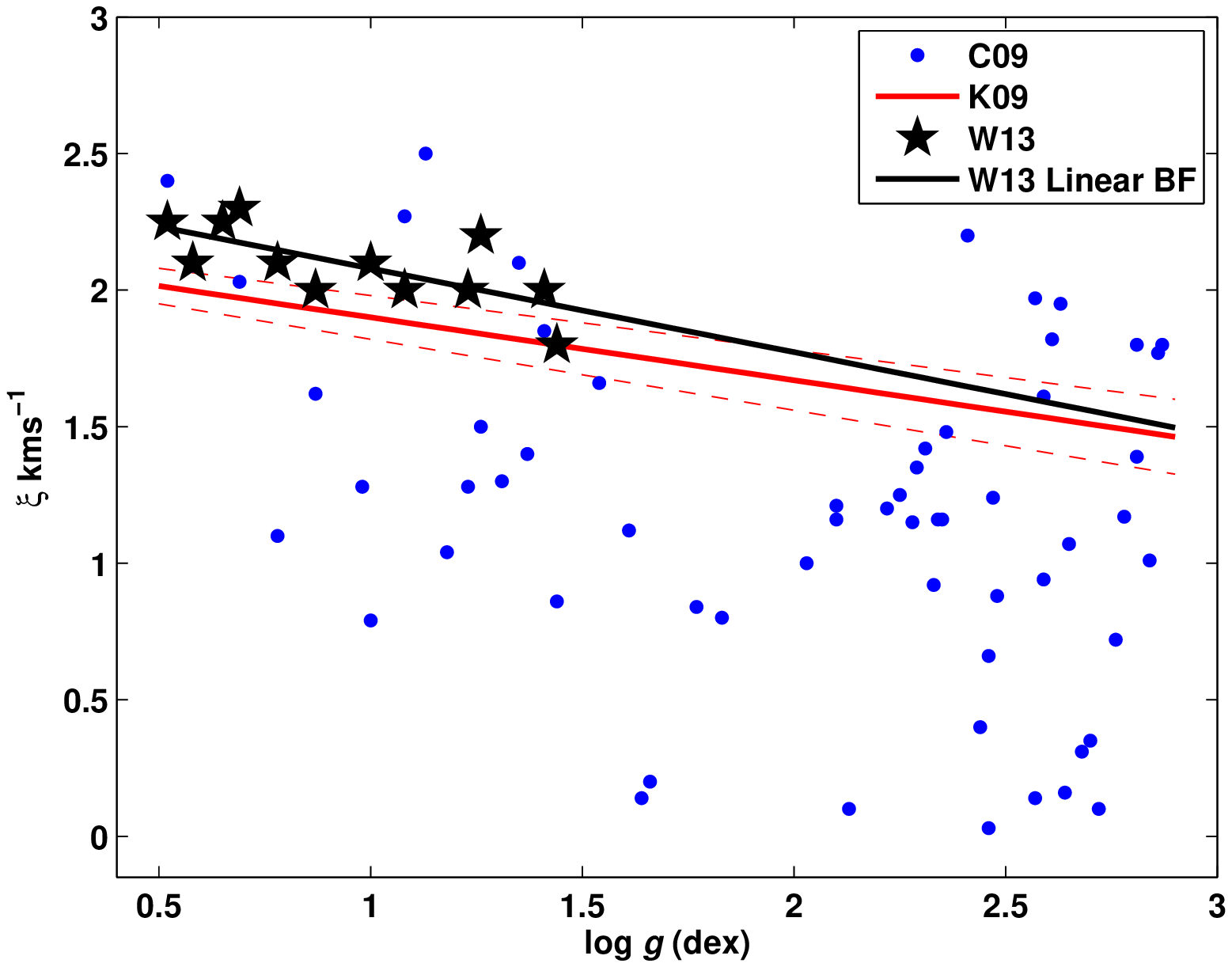}
\caption{Comparison of C09 $\xi$ and $\log g$ values with the $\xi$-$\log g$ relation from K09, the $\xi$ values determined here for the sample of 12 stars and the resulting $\xi$-$\log g$ relation that was derived.}\label{fig:vtregress}
\end{minipage}
\end{figure*}

Table~\ref{tab:meaEW_feifeii} lists the C09 parameters for 13 stars of which 12 had a sufficient sample of Fe lines that could be used to determine the microturbulence. The microturbulence determined in this analysis ($\xi_{W13}$) with the mean [Fe~I/H], mean [Fe~II/H], associated standard deviations and number of lines measured are also listed. For four of the spectra the equivalent widths of Fe~I lines measured in C09 for the corresponding stars were also used and the number of lines for each are also indicated. Hence for each of these 12 stars the equivalent widths of all possible Fe~I and Fe~II lines were measured and abundances determined using the curve of growth analysis routine, {\it abfind}, in the one-dimensional, local thermodynamic equilibrium (LTE) radiative transfer code, MOOG \citep{MOOG}. Stellar models for each star were generated at the C09 stellar parameters using the Kurucz 1995 stellar model grid and the ATLAS9\footnote{http://kurucz.harvard.edu/programs.html} model atmosphere 
calculation programme.

Eight of this sample of stars had also been analysed in previous studies, as well a ninth star for which Fe~I lines could not be measured here. The papers associated with each star are also listed in Table~\ref{tab:meaEW_feifeii}.

Figure~\ref{fig:fe1fe2comp} compares the mean [Fe~I/H] and mean [Fe~II/H] values measured here with those measured in C09 for the 12 stars, although in C09 Fe~II lines could only be measured for 6 of the current sample. The errorbars on the W13 measurements of Fe~II lines come from the uncertainty on the measurement of the Fe~I line in the corresponding star, as the dispersion around the Fe~I lines is a fair representation of the observational scatter for the measurement of weak lines in these spectra. Thus Figure~\ref{fig:fe1fe2comp} shows that there is agreement within 1~$\sigma$ between C09 and W13 for [Fe~I/H] and [Fe~II/H].

Based on the $\xi_{W13}$ derived for these 12 stars a $\xi$-$\log g$ relation was determined as the linear best fit to the 12 data points. This relation was then used to provide a $\xi_{W13}$ value for the remaining stars in the sample calculated from their $\log g$ value. Figure~\ref{fig:vtregress} shows the $\xi$ and $\log g$ values as determined in C09 and in this study (W13) with the $\xi$-$\log g$ relation and uncertainty limits determined in \citet{Kirby2009} (K09). The linear best fit to the $\xi_{W13}$ values is also shown, extended across the range of $\log g$ values for this sample. The equations for the K09 relation and relation derived here are as follows:
\begin{eqnarray}
\xi_{K09} &=& (2.13 \pm 0.05) - (0.23 \pm 0.03) \log g, \\
\xi_{W13} &=& 2.386-0.3067 \log g.
\end{eqnarray}
This sample is tailored specifically to these mono-metallicity stars while the K09 relation was derived from a sample of stars covering a range of metallicities. However the relations are in reasonably good agreement and in the expected range for giant stars.


\begin{figure*}[!t]
\begin{minipage}{185mm}  
\includegraphics[width=200mm]{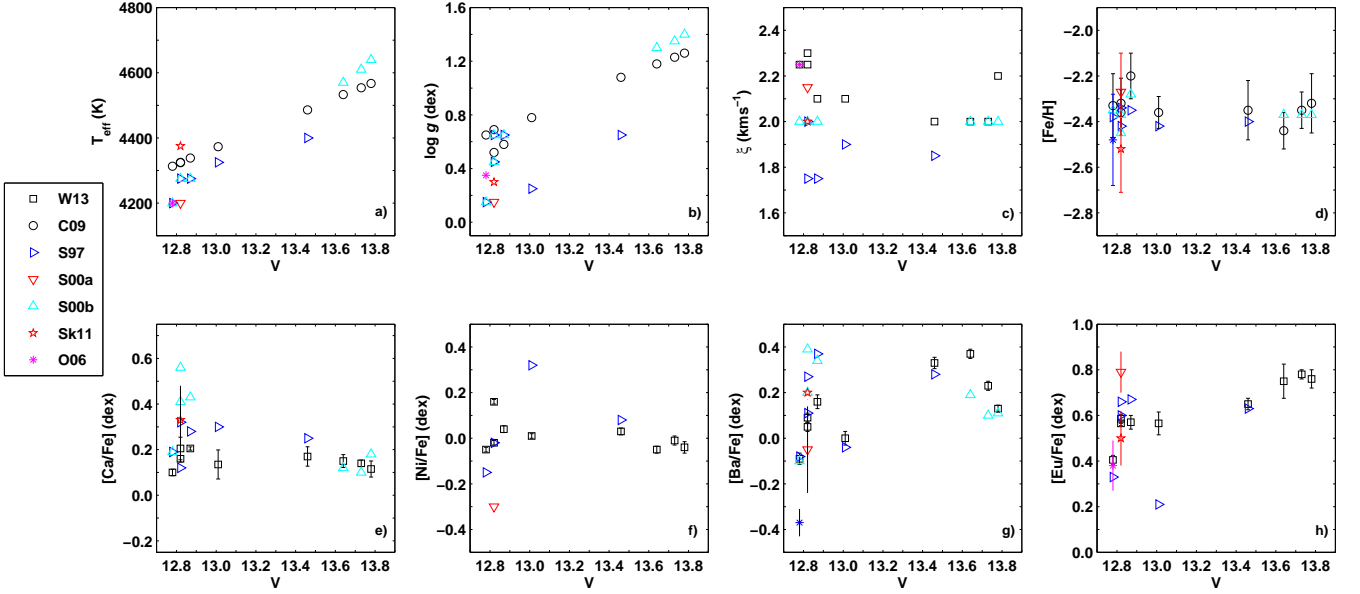}
\caption{Stellar parameters and chemical abundances against V magnitude for the 9 comparison stars in the M15 sample. The values from C09, W13, S97, S00a, S00b, O06 and Sk11 are shown: a)$T_{\textrm{eff}}$; b) $\log g$; c) $\xi$; d) [Fe/H]; e) [Ca/H]; f) [Ni/H]; g) [Ba/H]; and g) [Eu/H].}\label{fig:comppapers}
\end{minipage}
\end{figure*}

\section{Determination of Stellar Abundances}\label{sec:stellarabund}
As mentioned above the iron abundances based on Fe~I and Fe~II lines were determined using equivalent widths and curve-of-growth analysis in MOOG. For the determination of the abundances for Ca, Ni, Ba, La and Eu the techniques of spectrum synthesis were employed, again using MOOG and the Kurucz 1995 stellar models interpolated to the C09 parameters but now with the $\xi_{W13}$ values. The most recent laboratory $\log gf$s were obtained, as listed in Table~\ref{tab:linelist} and the hyperfine structure for the Ba~II, La~II and Eu~II were taken into account as per the decomposition in Table~\ref{tab:hfs} based on the most recent atomic information \citep{Rutten1978,Davidson1992,Lawler2001a,Lawler2001b}.

\subsection{Comparison Stars}
As shown in Table~\ref{tab:meaEW_feifeii} there are nine stars in this sample that have been analysed in previous studies. Figure~\ref{fig:comppapers} shows the range in values obtained for these stars for their stellar parameters and individual element abundances.

The stellar parameters of $T_{\textrm{eff}}$ are mostly within 50~K in terms of agreement between the studies (Figure~\ref{fig:comppapers}a). However for four of the stars the spread in $\log g$ is on the order of 0.5~dex  (Figure~\ref{fig:comppapers}b) and $\xi$ is particularly spread out ($\sim$0.3~kms$^{-1}$) between the studies for the high luminosity, low $\log g$ stars (Figure~\ref{fig:comppapers}c), which is reflective of the difficulty of deriving parameters for these types of stars. These differences reflect the degree of variation in stellar parameters that can be derived for the same star using different observations, measurements and analysis techniques. For example the three hottest stars have parameters from C09 and S00b only. For S00b, if the stars were not from S97, then the parameters were inferred from photometry-$T_{\textrm{eff}}$, $\log g$ relations. Hence the C09 parameters and S00b parameters are in good agreement for $T_{\textrm{eff}}$ and $\log g$, compared to the next two coolest 
stars which have C09 and S97 parameters. S97 derived parameters spectroscopically and while the $T_{\textrm{eff}}$ agree reasonably well with C09 the $\log g$ do not. However the good agreement in [Fe/H] between the studies is seen in Figure~\ref{fig:comppapers}c. 

Figure~\ref{fig:comppapers}d to h compare the abundances for Ca, Ni, Ba and Eu derived here with the abundances from the other studies for the comparison stars. There is reasonable agreement between the studies. Ca in particular shows good agreement between the studies at a reasonably constant absolute abundance. Ni is constant in abundance for this study but the three comparisons to two other studies show some variation. The range in abundance values between the stars for Ba and Eu is obvious already for this small sample of stars. For the chemical abundances the disagreement between studies is greatest for the high luminosity, reflecting the uncertainties in stellar parameters and chemical abundance determination for stars of low $\log g$. There may also be uncertainties in the model atmospheres (eg. sphericity effects) and in non-local thermodynamic equilibrium effects associated with cool giants of low gravity.

\begin{table}[!h]
\caption{Variations in measured abundances with changes in stellar atmospheric parameters for the sample of Fe~I and Fe~II lines, by equivalent width measurement, and the other key lines used in this study.}
\centering
\begin{tabular}{cccc}
\hline\hline
 & $\Delta T_{\textrm{eff}}$ & $\Delta \log g$ & $\Delta \xi$ \\ 
 & +50~K & +0.5~dex & --0.25~kms$^{-1}$  \\ 
\hline
$\Delta$[Fe~I/H] & 0.10 & --0.06 & 0.06 \\ 
$\Delta$[Fe~II/H] & -0.02 & 0.18 & 0.01 \\ 
$\Delta$[Ca I/Fe](6439.07) & 0.10 & --0.08 & 0.14 \\ 
$\Delta$[Ca I/Fe](6471.66) & 0.06 & --0.06 & 0.02 \\ 
$\Delta$[Ni I/Fe] & 0.10 & --0.03 & 0.07 \\ 
$\Delta$[Ba II/Fe] & 0.05 & 0.17 & 0.23 \\ 
$\Delta$[Eu II/Fe](6437.64) & 0.00 & 0.16 & 0.08 \\ 
$\Delta$[Eu II/Fe](6645.13) & 0.00 & 0.17 & 0.00 \\ 
$\Delta$[La II/Fe] & 0.03 & 0.18 & 0.01 \\ 
\hline
\end{tabular}
\label{tab:param_error}
\end{table}

Table~\ref{tab:param_error} gives the expected variations in the derived abundances for a $\Delta T_{\textrm{eff}}$=50~K, $\Delta \log g$=+0.5~dex and $\Delta \xi$=--0.25~kms$^{-1}$ for the light and heavy element lines being analysed in this study. The neutral species show greater sensitivity to changes in $T_{\textrm{eff}}$, while the ionised species show greater sensitivity to changes in $\log g$ as expected. The Ca~I line at 6471.66~\AA\ and the Ba~II line both show a sensitivity to changes in $\xi$, in particular the Ba II line which varies by 0.23~dex, as is expected for strong lines.

The possible effects of departures from LTE on the Ba spectral feature at 6496~\AA\ have been previously examined by \citet{Short2006} and \citet{Andrievsky2009}. \citet{Short2006} analysed the non-LTE sensitivity of the Ba II 6496 transition for a variety of different stellar atmospheric model and line transfer combinations in a typical halo giant (with T$_{eff}$~=~4800~K and $\log g$~=~1.5).   By and large, they found a metallicity dependence of the non-LTE effect with line strength (explicitly, a non-LTE weakening of transitions for the most metal-deficient models and conversely, a non-LTE strengthening of lines for the least metal-poor models).  Yet, in particular for the 6496 transition, when the data for full non-LTE model/transfer are compared to those for LTE model/transfer, the magnitude of the change was relatively small on the resultant Ba abundance (with $|\log $W$_{\lambda}$(NLTEFull)$ - \log$W$_{\lambda}$(LTE)$|$ $\leq$ 0.03).

On the other hand, \citet{Andrievsky2009} employed a single LTE atmospheric model with non-LTE radiative transfer to study the non-LTE effect on Ba II lines as a function of both metallicity and temperature in extremely metal-poor giants.  Their results mostly confirm those of \citet{Short2006} (a correlation between metallicity and non-LTE departure was established with the largest effect seen for warm, metal-poor stars). Specifically for the Ba II 6496~\AA\ feature, they found a maximum negative correction factor (NLTE-LTE) of $\approx$--0.1 for a giant with [Fe/H]~=--2.0 and T$_{eff}$~=~4700~K and a maximum positive (NLTE--LTE) correction of $\approx$--0.13 for a giant with [Fe/H~]=--2.5 and T$_{eff}$~=~5500~K.  \citet{Andrievsky2009} note, however, that the correction factor cannot be used directly to modify the Ba abundance (and rather, the comparison of non-LTE to LTE line profiles is requisite).  Consequently, in consideration of these findings as well as those of the current study, the sensitivity of 
the 6496~\AA\ transition to departures from LTE appears to be a minor and secondary source of error for the M15 target stars.

\section{Chemical Abundances in the M15 Sample}\label{sec:stellarabundanalysis}
The wavelength range and resolution of the spectra observed for this study allowed for the analysis of two Ca~I lines, one Ni~I line, one Ba~II line, one La~II line and two Eu~II lines (see Table~\ref{tab:linelist}). Abundances from these lines were determined using the spectrum synthesis routine, {\it synth}, in MOOG. For each of the stars in the sample Table~\ref{tab:chemabund} lists the chemical abundances and the associated fitting errors, which were determined as the greatest change in abundance for possible fits to the spectral feature. Also listed are the C09 parameters and chemical abundances, $\xi_{W13}$ and the S/N measured here.

In order to extract further information from the current M15 sample, groups of stars were identified with similar stellar parameters and low Ba abundances. Due to the noise limitations, the key spectral features of Eu and La could not be measured confidently for these individual spectra. However they could be combined into two summed spectra such that the noise was sufficiently reduced to estimate the `summed' heavy element abundances. Table~\ref{tab:abund_sumspe} gives the mean parameters for each set of spectra including the mean [Ba/Fe] for each set, the [Ba/Fe] derived from the summed spectra, the upper limits on [Eu/Fe] for both summed spectra, and upper limit for [La/Fe] for just one of the summed spectra. 

\begin{table}[!h]
\centering
\caption{Average stellar parameters of two sets of summed spectra and the resulting heavy element abundances.}
\begin{tabular}{c|cc}
\hline\hline
 & A & B \\ 
\hline
$\langle$S/N$\rangle$ & 121$\pm$25 & 70$\pm$11 \\ 
$\langle T_{\textrm{eff}} \rangle$ & 4949$\pm$116 & 5216$\pm$43 \\ 
$\langle \log g \rangle$ & 2.1$\pm$0.3 & 2.6$\pm$0.1 \\ 
$\langle$[Fe/H]$\rangle$ & --2.3$\pm$0.1 & --2.4$\pm$0.1 \\ 
$\langle \xi_{W13} \rangle$ & 1.7 & 1.6 \\ 
$\langle[$Ba$/$Fe$]\rangle$ & --0.16$\pm$0.05 & --0.33$\pm$0.17 \\ 
$[$Ba$/$Fe$]$ & --0.15$\pm$0.08 & --0.39$\pm$0.15 \\ 
$[$Eu$/$Fe$]$ & $<$0.42 & $<$0.41 \\ 
$[$La$/$Fe$]$ & $<$0.37 & -  \\ 
 &  &  \\ 
C09 ID & \multicolumn{2}{l}{A= 18815, 42362, 27889} \\ 
 & \multicolumn{2}{l}{B= 35961, 23216, 23153, 8927} \\ 
\hline
\end{tabular}
\label{tab:abund_sumspe}
\end{table}

\begin{figure*}[!t]
\centering
\begin{minipage}{180mm}  
\centering
\includegraphics[width=180mm]{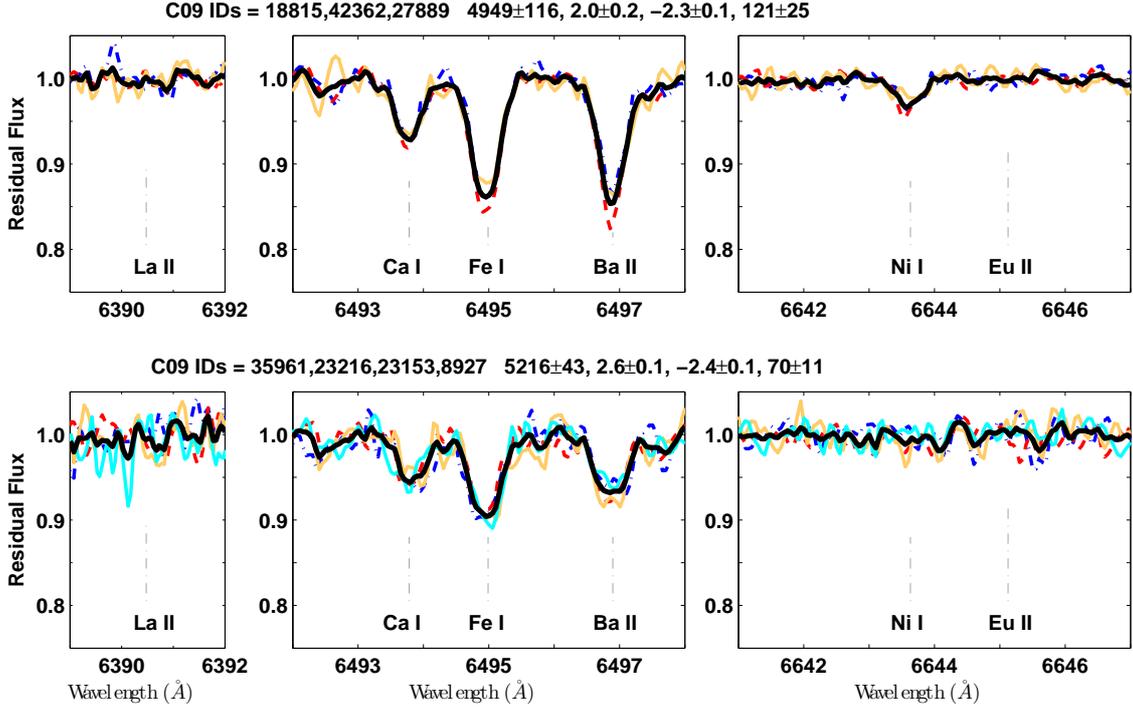}
\caption{The three component spectra (blue, red, orange) and resulting summed spectrum (black) for A in the top panels, and the four component spectra (blue, red, orange, cyan) and resulting summed spectrum (black) for B in the bottom panels, in the regions of the La II, Ba II and Eu II spectral lines respectively. The C09 IDs for each set of spectra and the resulting mean stellar parameters ($T_{eff}$, $\log g$, [Fe/H]) and S/N for the final summed spectrum for each of A and B are given.}\label{fig:sumspe}
\end{minipage}
\end{figure*}

Figure~\ref{fig:sumspe} shows the component spectra and the resulting summed spectra of A and B in the regions about the three key heavy element spectral features. The component spectra of B are much noisier than A as expected due to being fainter (hotter) on the RGB and due to it being an inherently weak line, the La spectral line profile was still too weak to be detected above the noise. Hence no upper limit on the La abundance could be extracted for summed spectrum B. The goal of the summed spectra analysis was to investigate the Eu abundance at low Ba abundances. This will be discussed further in the analysis of the results.

The analysis of this sample of stars in M15 must be considered in light of the results of the previous studies. The mean values for the metallicity, Ba, Eu and La abundances for each key of M15 study are given in Table~\ref{tab:means4cluster}, where the C09 metallicity represents the mean metallicity for this sample.

\begin{table*}[!t]
\caption{The number of stars in the sample, the mean and standard deviations for [Fe/H], [Ba/H], [La/H] and [Eu/H] for each key study.}
\begin{center}
\begin{tabular}{cccccc}
\hline
 &  N$_{star}$ &$<$[Fe/H]$>\pm\sigma$ & $<$[Ba/H]$>\pm\sigma$ & $<$[La/H]$>\pm\sigma$ & $<$[Eu/H]$>\pm\sigma$ \\ 
\hline\hline
W13   & 63 & -                 & --2.20$\pm$0.26   & --2.05$\pm$0.15 & --1.72$\pm$0.15 \\ 
Sk11* & 3  & --2.55$\pm$0.02   & --2.10$\pm$0.12   & --2.07$\pm$0.23 & --1.72$\pm$0.29 \\ 
DO10  & 57 & -                 & --2.152$\pm$0.412 & -               & - \\ 
C09   & 84 & --2.341$\pm$0.007 & -                 & -               & -  \\ 
O06   & 7  & --2.40$\pm$0.05   & --2.56$\pm$0.18   & --2.46$\pm$0.25 & --1.85$\pm$0.24 \\ 
S00b  & 31 & --2.37$\pm$0.05   & --2.25$\pm$0.21   & -               & -  \\ 
S00a  & 3  & --2.28$\pm$0.05   & --2.28$\pm$0.19   & -               & --1.41$\pm$0.31 \\ 
S97   & 18 & --2.40$\pm$0.04   & --2.30$\pm$0.21   & -               & --1.91$\pm$0.21 \\ 
\multicolumn{ 6}{l}{* RGB stars only} \\ 
\hline
\end{tabular}
\end{center}
\label{tab:means4cluster}
\end{table*}

What is noticeable is that the mean Ba abundance derived in this study is some 0.2~dex greater than for the majority of the other studies. Looking particularly at S97 and S00b, which have sizeable samples likely to sample the large range in Ba abundances, there is an offset of +0.12~dex to the mean Ba abundance of S97 and an offset of +0.17 to the mean Ba abundance of S00b. Both studies have dispersions in Ba abundance which are comparable to the spread in Ba measured for this sample. Hence the large range of Ba values for M15 is confirmed in this study but the absolute abundances appear offset to previous studies. However it should again be noted that the 63 stars analysed here are the largest sample by far of M15 stars for which Ba abundances have been determined, the next largest sample for Ba abundances being S00b with 30 stars. This sample is also complete along the RGB based on the selection criteria for C09.

The comparison to DO10 is unique in that both D010 and this study are based on slightly different subsamples of C09. Interestingly the mean Ba abundances of these two studies are in reasonable agreement although the spread reported in D010 is greater. As was discussed in Section~\ref{sec:stellarparams}, the $\xi$ values determined in C09 resulted in a trend of Ba abundance with $\xi$. Hence the mean Ba abundance from D010 must be considered in light of this, as no adjustment to the $\xi$ values was made. The equivalent width of this alternate subsample of C09 were provided (D'Orazi, private communication) in order to make a better comparison of the Ba abundances using the $\log g - \xi$ relation derived here. 

Considering the mean abundances for La and Eu given in Table~\ref{tab:means4cluster}, the mean La abundance agrees well with Sk11, but O06 is overestimated in comparison. On the other hand the mean Eu abundance for this study lies within the range of values determined for the other studies. 

The offset in Ba abundance between the studies can be attributable to differences in the measurement techniques and analysis methods used to derive the stellar parameters and abundances. As a key example, the sample of 18 stars in S97 were analysed by the measurement of equivalents widths which were then used in a curve-of-growth analysis to determined the stellar parameters and abundances, whereas for C09 the stellar parameters were determined via photometry and fiducial relations configured for M15. In particular, the Ba abundance determined in S97 was based on three Ba lines for which equivalent widths were measured. Two of the lines (6141.73~\AA, 6496.91~\AA) were typically measured to have equivalent widths at $\sim$130~m\AA, while the third (5853.69~\AA) at $\sim$80~m\AA.

There is a similar difference in analysis method with DO10 for which equivalent width techniques were also employed in the analysis of the single Ba line at 6141.73~\AA.

As S97 reports the equivalent widths that were measured for the 18 stars, and the equivalent widths from the DO10 analysis were also provided privately by the authors, we pursued extending the current analysis to include the stars from these papers. 

\begin{table*}[!th]
\centering
\caption{Ba equivalent widths (EW) measurements from S97 and DO10 and the corresponding [Ba/H] and [Eu/H] derived from these measurements based on the C09 parameters and W13 $\xi$ for the cross-over stars between W13, S97 and DO10. The W13 abundances are based on spectrum synthesis.}
\begin{tabular}{cc|ccc|cccc|cc}
\hline
&  &   \multicolumn{ 2}{l}{Ba EW (m\AA)} &  & [Ba/H] &  &  &  & [Eu/H] &  \\ 
C09 ID & S97 ID & {\scriptsize S97 6496} & {\scriptsize S97 6141} & {\scriptsize DO10 6141} & {\scriptsize W13 6496} & {\scriptsize S97$_{W13}$ 6496} & {\scriptsize S97$_{W13}$ 6141} & {\scriptsize DO10$_{W13}$ 6141} & {\scriptsize W13 6645} & {\scriptsize S97$_{W13}$ 6645} \\ 
\hline\hline
43788 & K825 & 136 & 137 & 137 & --2.27 & --2.52 & --2.63 & --2.63 & --1.85 & --1.84 \\ 
40825 & K386 & 129 & 133 & 140 & --2.42 & --2.65 & --2.72 & --2.61 & --1.94 & --1.87 \\ 
4099 & K341 & 146 & 146 & - & --2.27 & --2.42 & --2.53 & - & --1.67 & --1.66 \\ 
31914 & K757 & 151 & 147 & 154 & --2.04 & --2.23 & --2.28 & --2.39 & --1.67 & --1.72 \\ 
41287 & K702 & 130 & 131 & - & --2.36 & --2.46 & --2.26 & - & --1.90 & --2.07 \\ 
3137 & K387 & 143 & 139 & 133 & --2.02 & --1.96 & --2.17 & --2.28 & --1.73 & --1.65 \\ 
\hline
\end{tabular}
\label{tab:s97c09_crossbaeu}
\end{table*}

As outlined in Table~\ref{tab:meaEW_feifeii} six of the stars in S97 are already part of the current sample. Four of these stars are also part of the DO10 sample. In total there are 38 stars in common between DO10 and W13. Using the corresponding C09 parameters with the modified $\xi$ value, the Ba (6141.73~\AA, 6496.91~\AA) and Eu (6645~\AA) abundances were re-determined using the S97 equivalent widths and the DO10 equivalent widths for the Ba (6141.73~\AA) abundance. 

Simple low order polynomial relations were derived for $T_{\textrm{eff}}$ and $\log g$ as functions of V based on the entire C09 sample. These functions were then used to calculate the corresponding $T_{\textrm{eff}}$ and $\log g$, and from $\log g$ the $\xi$, for each of the 12 S97 stars and 19 DO10 stars. The re-analysis of the DO10 and S97 samples using C09/W13 parameters and the resulting abundances are referred to as S97$_{W13}$ and DO10$_{W13}$ for the remainder of the paper. 

\begin{figure}[!t]
\centering
\begin{minipage}{90mm}  
\hspace{-0.3cm}
\includegraphics[width=95mm]{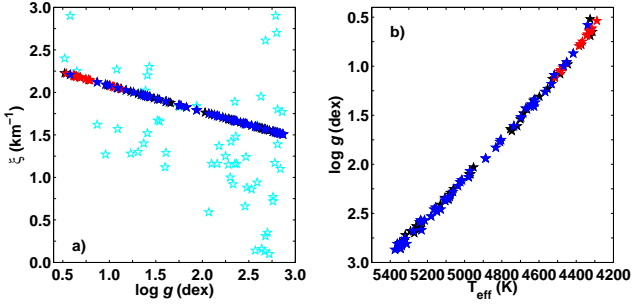}
\caption{C09 stellar parameters and W13 $\xi$ attributed to the S97 and DO10 samples. a) $\xi$ against $\log g$ for C09 (black), S97$_{W13}$ (red) and DO10$_{W13}$ (blue) and DO10 (cyan). b) HR diagram for C09, S97$_{W13}$ and DO10$_{W13}$.}\label{fig:s97c09params}
\end{minipage}
\end{figure}

Figure~\ref{fig:s97c09params}a shows $\xi$ compared with $\log g$ for the C09 (black), DO10$_{W13}$ (blue) and S97$_{W13}$ (red). The original $\xi$ values used in DO10 (cyan) are shown for comparison. Figure~\ref{fig:s97c09params}b is corresponding HR diagram for C09, DO10$_{W13}$ and S97$_{W13}$. 

Using these parameters, and assuming a metallicity of [Fe/H]~=--2.341~dex (C09), stellar models were interpolated using ATLAS9 as for the stellar models for the W13 sample. Then, using the equivalents widths from each study, the abundances for Ba and Eu  were derived using the {\it blends} routine in MOOG taking account of hyperfine structure using the same linelists as used in the spectrum synthesis. 

The cross-over stars between the three studies were investigated to look for potential systematic trends between the samples. Table~\ref{tab:s97c09_crossbaeu} lists the six stars in common between W13, S97 and DO10 with their respective IDs, the Ba equivalent widths from S97 and DO10, and the resulting derived abundances where possible (S97$_{W13}$, DO10$_{W13}$). While two Eu lines were used in the spectrum synthesis analysis in this paper only the abundance determined for the Eu line in common with S97 (6645~\AA) was used to determine the bias between the techniques.

\begin{figure}[!t]
\centering
\begin{minipage}{90mm}  
\hspace{-0.3cm}
\includegraphics[width=95mm]{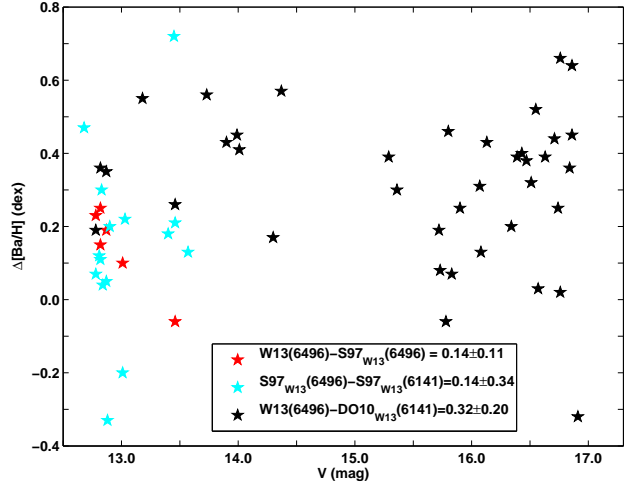}
\caption{Comparison of the [Ba/H] derived for the stars in common between W13, S97$_{W13}$ and DO10$_{W13}$ as per the legend against V magnitude. Biases and dispersions between each sample are listed.}\label{fig:bias_w12s97do10}
\end{minipage}
\end{figure}

Two effects can be seen here and are illustrated in Figure~\ref{fig:bias_w12s97do10}, which shows the $\Delta$[Ba/H] between the cross-matched stars of W13, S97$_{W13}$ and DO10$_{W13}$. Although for just a sample of six stars there is a bias of $\sim$0.14~dex between the [Ba/H] abundances derived from the specturm synthesis in W13 compared with the S97$_{W13}$ [Ba/H] abundances derived from the equivalent width measurements. This first effects shows that even for the same lines differences in measurements and analysis techniques lead to a difference in derived abundances.

The second effect is illustrated by the comparison of the S97$_{W13}$ [Ba/H] values derived from the 6496~\AA\ line compared with the 6141~\AA\ line. There is also a bias of $\sim$0.14~dex combined with a significant scatter ($\sim$0.34). Although great care was taken here deriving the $\xi$ values this discrepency between the two strong lines measured in the same study may show the derived relation is still not optimised. 

Hence not only differences between measurement techniques and analysis must be taken account of, but also differences in the abundances derived by the same method but of two different Ba lines. This is reflected in the comparison of W13 to the DO10$_{W13}$ [Ba/H], which is a comparison of spectrum synthesis to equivalent width determinations as well as a comparison between the 6496~\AA\ and 6141~\AA\ lines. The bias between these two samples is $\sim$0.32$\pm$0.20~dex, which equates to the combination of the biases produced by these two effects. 

\begin{figure*}[!t]
\hspace{-1cm}
\begin{minipage}{90mm}  
\centering
\includegraphics[width=105mm]{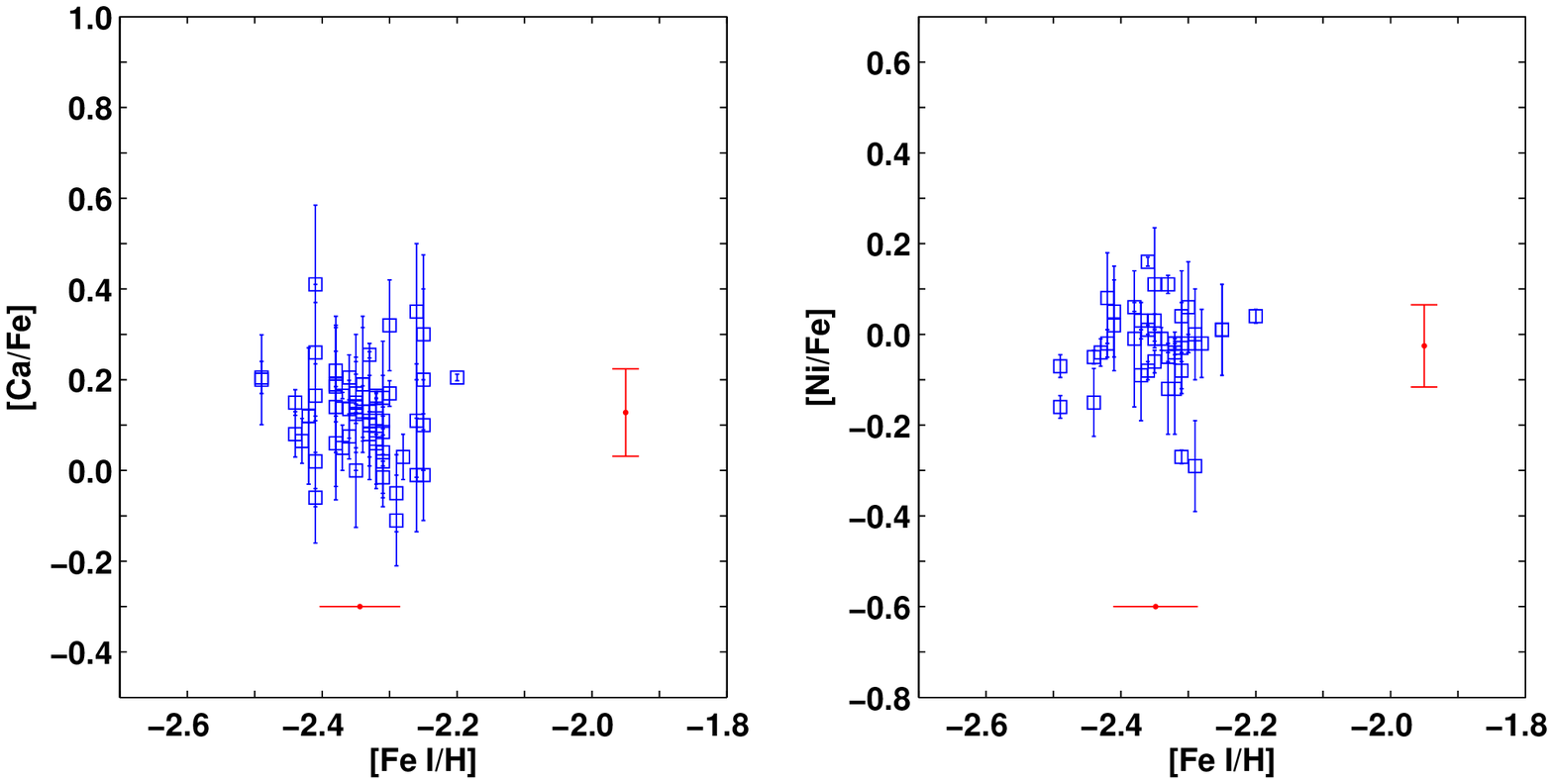}
\end{minipage}
\hspace{0.6cm}
\begin{minipage}{90mm}
\includegraphics[width=105mm]{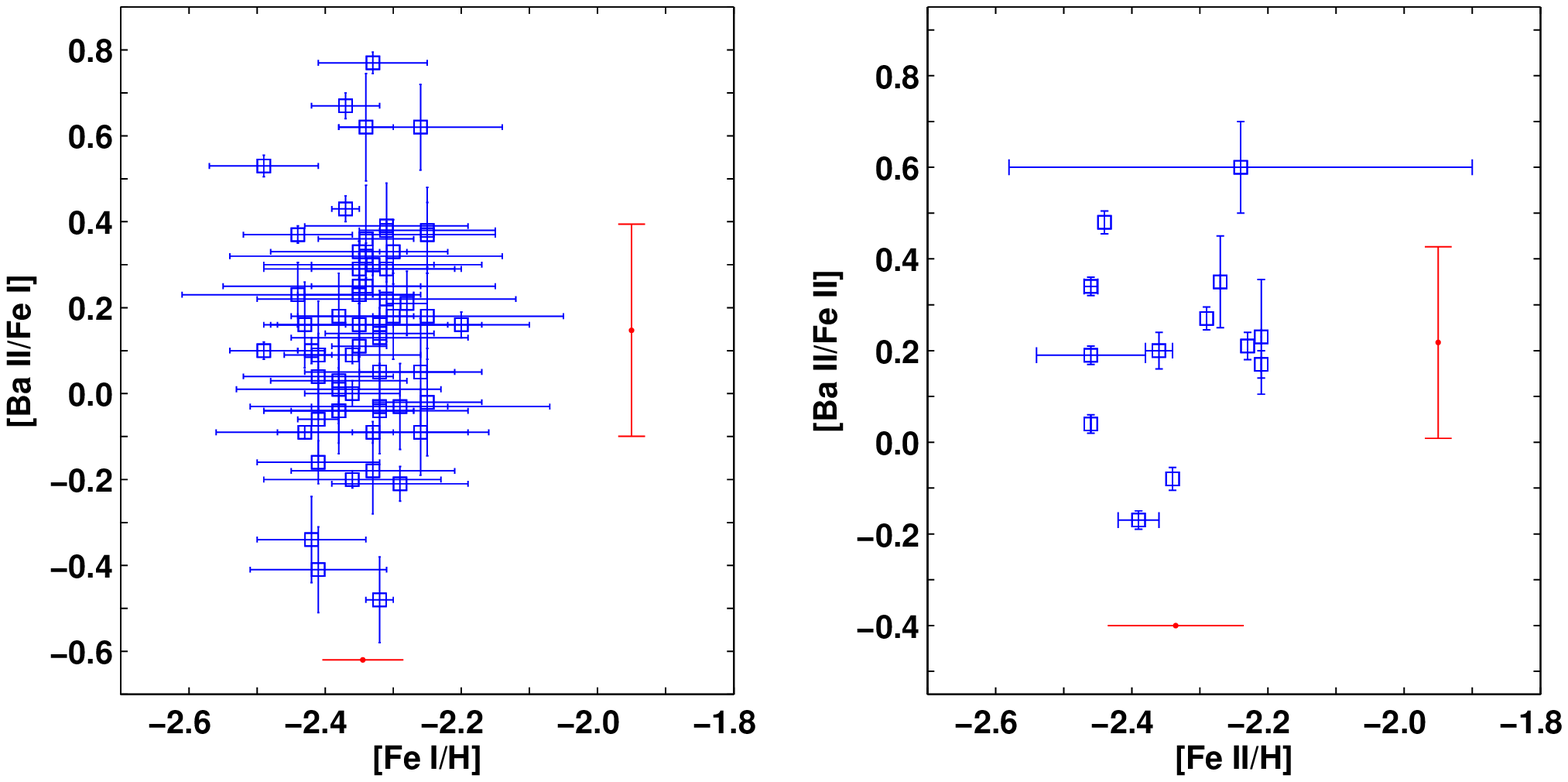}
\end{minipage}
\caption{Comparisons between chemical abundances measured here with Fe from C09 for this sample: a) [Ca I/Fe] vs [Fe I/H]; b)[Ni I/Fe] vs [Fe I/H]; c) [Ba II/H] vs [Fe I/H]; and d) [Ba II/H] against [Fe II/H]. Errors for Ca, Ni and Ba reflect the fitting errors of the spectrum synthesis. Mean values and uncertainties for each sample are shown in red.}\label{fig:canionafe}
\end{figure*}

\begin{table}[htbp]
\centering
\caption{[Ba/H] and [Eu/H] derived for the 12 stars of S97 and the 19 stars of DO10 that are not part of the W13 sample. For S97$_{W13}$ only $\lambda$(Ba)~=~6496~\AA\ and $\lambda$(Eu)~=~6645~\AA\ are used. The C09 stellar atmosphere parameters (with [M/H]~=--2.341~dex for S97$_{W13}$) and $\xi_{W13 }$ are listed.}
\begin{tabular}{cccccccc}
\hline
S97 ID & V & $T_{\textrm{eff}}$ & $\log g$ & [M/H] & $\xi$ & [Ba/H] & [Eu/H] \\ 
\hline\hline
K479 & 12.68 & 4288 & 0.54 & --2.341 & 2.2 & --1.99 & --1.84 \\ 
K634 & 12.81 & 4321 & 0.63 & --2.341 & 2.2 & --2.56 & --1.96 \\ 
K583 & 12.83 & 4327 & 0.64 & --2.341 & 2.2 & --2.55 & --2.12 \\ 
K490 & 12.84 & 4329 & 0.65 & --2.341 & 2.2 & --2.87 & --2.06 \\ 
K853 & 12.88 & 4339 & 0.67 & --2.341 & 2.2 & --2.65 & --1.99 \\ 
K462 & 12.90 & 4344 & 0.69 & --2.341 & 2.2 & --2.31 & --1.49 \\ 
K431 & 13.03 & 4377 & 0.77 & --2.341 & 2.1 & --2.46 & --1.95 \\ 
K144 & 13.06 & 4385 & 0.79 & --2.341 & 2.1 & --2.26 & --1.70 \\ 
K1040 & 13.40 & 4471 & 1.01 & --2.341 & 2.1 & --2.19 & --1.68 \\ 
K969 & 13.45 & 4484 & 1.04 & --2.341 & 2.1 & --1.53 & --1.95 \\ 
K169 & 13.47 & 4489 & 1.06 & --2.341 & 2.1 & --2.26 & --1.69 \\ 
K146 & 13.57 & 4514 & 1.12 & --2.341 & 2.0 & --2.53 & --2.07 \\ 
 &  &  &  &  &  &  &  \\ 
C09 ID &  &  &  &  &  &  &  \\ 
\hline\hline
34332 & 13.32 & 4451 & 0.96 & --2.41 & 2.1 & --2.42 & - \\ 
14574 & 13.58 & 4518 & 1.09 & --2.34 & 2.1 & --2.65 & - \\ 
25247 & 14.00 & 4623 & 1.34 & --2.34 & 2.0 & --2.35 & - \\ 
34561 & 14.07 & 4641 & 1.39 & --2.27 & 2.0 & --2.15 & - \\ 
29401 & 14.09 & 4646 & 1.41 & --2.34 & 2.0 & --2.78 & - \\ 
37415 & 14.19 & 4671 & 1.43 & --2.28 & 1.9 & --2.17 & - \\ 
9753 & 14.29 & 4697 & 1.52 & --2.40 & 1.9 & --2.73 & - \\ 
18913 & 14.44 & 4735 & 1.61 & --2.46 & 1.9 & --3.06 & - \\ 
31313 & 14.72 & 4805 & 1.75 & --2.28 & 1.8 & --2.08 & - \\ 
33507 & 15.04 & 4887 & 1.94 & --2.33 & 1.8 & --2.21 & - \\ 
41437 & 15.37 & 4971 & 2.07 & --2.29 & 1.8 & --2.68 & - \\ 
31956 & 15.54 & 5015 & 2.16 & --2.34 & 1.7 & --2.35 & - \\ 
30535 & 15.58 & 5024 & 2.18 & --2.24 & 1.7 & --2.07 & - \\ 
41279 & 15.82 & 5087 & 2.35 & --2.33 & 1.7 & --2.15 & - \\ 
28350 & 15.90 & 5105 & 2.37 & --2.37 & 1.7 & --2.86 & - \\ 
38382 & 16.19 & 5180 & 2.53 & --2.20 & 1.6 & --2.20 & - \\ 
24706 & 16.40 & 5232 & 2.67 & --2.42 & 1.6 & --3.08 & - \\ 
36971 & 16.75 & 5321 & 2.76 & --2.26 & 1.5 & --2.39 & - \\ 
20994 & 16.82 & 5338 & 2.79 & --2.28 & 1.5 & --2.51 & - \\ 
\hline
\end{tabular}
\label{tab:s97c09_EWbaeu}
\end{table}

In addition, while there is no particular trend of $\Delta$[Ba/H] with V, the brighter stars show a clear offset but the fainter stars have a much greater dispersion and less systematic offset. At the faint end the noise is greater, particularly shown by the greater fitting errors measured for W13, which can for example affect the placement of the continuum and so the derived abundance. 

Also there are the issues between measurement and analysis techniques, and that it is a comparison between two different Ba lines. The comparison between the S97 measurements shows also a large scatter in values. As much as great care was taken in deriving the $\xi$ values this discrepency between the two strong lines measured in the same study may show the derived relation is still not optimised. 

The bias for Eu calculated between W13 and S97$_{W13}$ is $\Delta$[Eu/H]=0.01$\pm$0.09~dex, which is negligible. The dispersion in the bias reflects the random errors as expected. This is not unexpected as the differences in the Ba abundances are most likely driven by the microturbulence, the effects of which are negligible on the weak spectral lines of Eu. The sensitivity of strong Ba features to microturbulence makes inter-study comparisons more problematic, even after considerations of differences in measurement and analysis techniques. The spread in the values of $\Delta$[Ba/H] are further evidence of the difficulty in obtaining consistency between studies.

The goal of this exploration was to see how to combine these samples. The offsets in themselves could be applied but the large scatter, particular between W13 and DO10$_{W13}$ for the faint stars, may reflect too many issues with the difference between the samples and techniques. It was decided that combining these samples potentially causes more problems then it solves but the samples can be analysed alongside each other for comparison as three distinct internally consistent analyses.

Table~\ref{tab:s97c09_EWbaeu} lists the 12 stars from S97 and the 19 stars from DO10, the stellar parameters calibrated to C09 that were used to generate the atmospheric models and the resulting [Ba/H] (from the 6496~\AA\ line only for S97) and [Eu/H] abundances (for S97 only).

\subsection{Abundance Distributions}
Ca abundances were determined for 62 of the 63 stars while Ni abundances were determined for 40 of the 63 stars. From the analysis in C09 23 of the 63 stars have derived O abundances and 49 have derived Na abundances (see Table~\ref{tab:m15_c09}). Figure~\ref{fig:canionafe} a. and b. show the distribution of [Ca/Fe] and [Ni/Fe] respectively compared to [Fe/H] for the M15 sample along with the sample mean values and standard deviations in red. 

Both Ca and Ni show a very tight distribution similar in magnitude to that of [Fe/H], implying that there is little star-to-star variation within the cluster of the abundances of these two elements. As described in full in C09, O and Na both have a large spread of abundance values between stars in M15. However, using the values from C09, no trend was found between Na nor O with any of the three heavy elements.

Figure~\ref{fig:canionafe}c. and d. compares the derived [Ba/H] values with the iron abundances determined from Fe~I and Fe~II lines in C09. There is no trend of Ba with Fe for either species and the larger spread of Ba compared with Fe is distinct.

\begin{figure*}[!t]
\begin{minipage}{185mm}  
\hspace{-2cm}
\includegraphics[width=220mm]{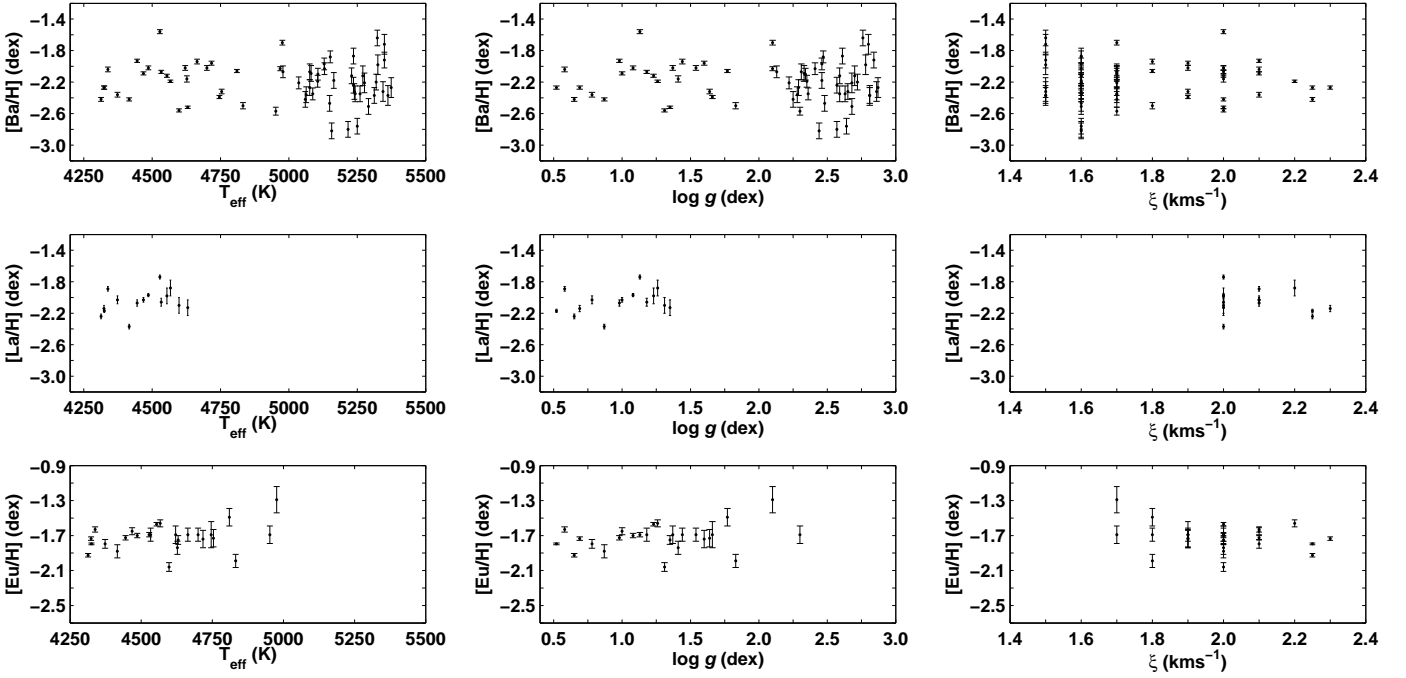}
\caption{Comparison of the [Ba/H], [La/H] and [Eu/H] values with the stellar parameters from C09. Errors on [Ba/H] and [La/H] are spectrum synthesis fitting errors. Errors on [Eu/H] are the spectrum synthesis fitting errors if just one line was measured, otherwise the standard deviation of the two Eu lines summed in quadrature with the fitting error.}\label{fig:baparams}
\end{minipage}
\end{figure*}

\subsection{Heavy Elements}
The primary goal of this study was to investigate Ba and Eu abundances in a large sample of M15 giant stars. While Ba could be measured for each of the 63 stars, Eu was measured for only 20 of the stars and La was measured for just 13 of the stars. Figure~\ref{fig:baparams} compares the derived [Ba/H], [La/H] and [Eu/H] values with the stellar parameters.

As discussed in Section~\ref{sec:stellarparams} the Ba~II spectral feature considered here is very sensitive to $\xi$ but there is clearly no trend of [Ba/H] with $\xi_{W13}$ despite the large spread in [Ba/H] values. This, as well as the uniform Ca abundances, for which one of the lines was also sensitive to $\xi$, confirms the appropriate derivation of the $\xi$--$\log g$ relation carried out for this sample.

The observations suffered from greater levels of noise in the lower part of the RGB (fainter stars) hence greater uncertainties are reported. The subsample for which La and Eu abundances could be determined from their respective weak spectral features were necessarily located high on the RGB (brightest stars).

\begin{figure}[!h]
\centering
\begin{minipage}{90mm}  
\hspace{-0.2cm}
\includegraphics[width=100mm]{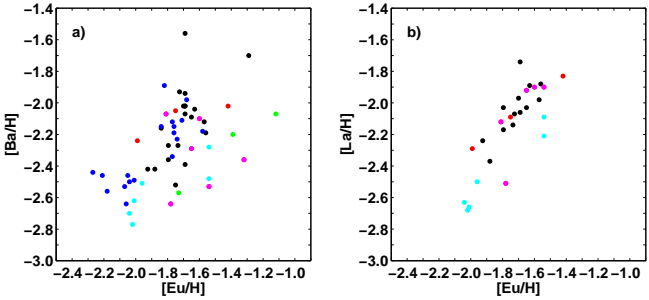}
\caption{a) [Ba/H] and b) [La/H] compared with [Eu/H] for the W13 (black) sample compared with S97 (blue), S00a (green), O06 (cyan), and Sk11 (red:RGB, pink:RHB) samples where available.}\label{fig:baeusnesob}
\end{minipage}
\end{figure}

Figure~\ref{fig:baeusnesob} combines the samples from the key studies of heavy elements in M15, including the W13 sample. The S97 sample is that as reported in that paper. Despite the range in measurement types and analysis techniques there is a clear trend of [Ba/H] with [Eu/H], and [La/H] with [Eu/H] even with the dispersion between the studies. However, as discussed in Section~\ref{sec:stellarabundanalysis}, there are distinct offsets between the studies which in a basic comparison, such as in Figure~\ref{fig:baeusnesob}, may blur any more refined trend in the GC itsself. Hence for the rest of this paper only the re-analysis of the S97 and DO10 samples are considered and are displayed separately to the W13 sample.

\begin{figure*}[!th]
\centering
\begin{minipage}{185mm}  
\hspace{-2cm}
\includegraphics[width=220mm]{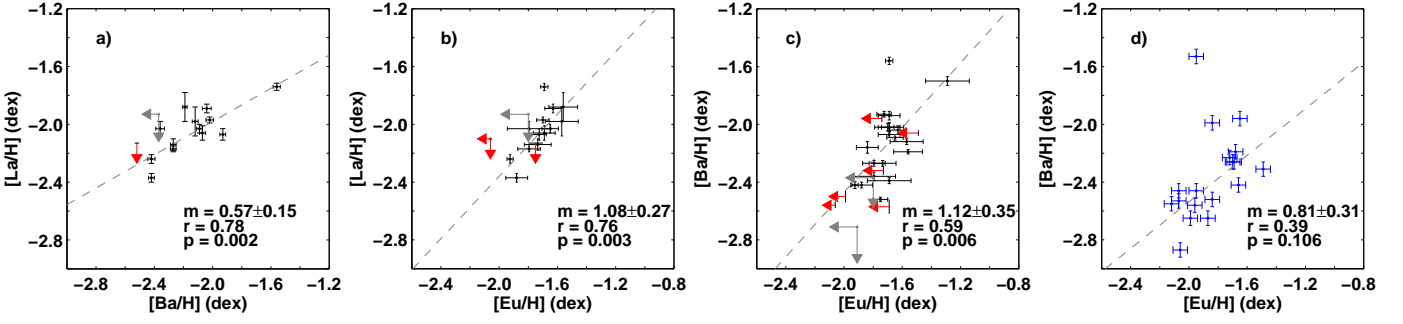}
\caption{Comparisons of [La/H] with [Ba/H] and [La/H] with [Eu/H] for the 13 W13 stars with La, Eu and Ba measurements (a \& b), and [Ba/H] with [Eu/H] for the 20 W13 stars with Ba and Eu measurements (c). The errorbars for La and Ba are the fitting errors, while for Eu the errorbars are the standard deviation if two lines were measured, otherwise the errorbars are the fitting errors. The line of best fit (grey dash) and its slope (m), as well as the correlation coefficient (r) and p-value (p) of each dataset is given. Red arrows are upper limits on the abundances for particular stars. Grey arrows are limits determined for the two sets of summed spectra. d) [Ba/H] vs [Eu/H] for the 18 S97$_{W13}$ stars.}\label{fig:labaeu}
\end{minipage}
\end{figure*}

Figure~\ref{fig:labaeu}a \& b compares [La/H] with [Ba/H] and [La/H] with [Eu/H] for the subsample of 13 W13 stars for which all three species could be measured, and Figure~\ref{fig:labaeu}c compares [Ba/H] with [Eu/H] for the 20 stars with both Eu and Ba abundances. Figure~\ref{fig:labaeu}d compares [Ba/H] with [Eu/H] for the 18 stars in S97. Included on each graph is the line of best fit to each dataset where the weighting of each point was defined as the inverse of the errors summed in quadrature. The slope (m), correlation coefficient (r) and p-value (p) for each comparison is given. Also included in red are upper limits in the cases for which a line could not be accurately measured but was still detectable. The limits derived from the two summed spectra are also included in grey.

For each comparison of the W13 samples the calculated p-value is less than 0.05 which indicates that the linear relation of these data is unlikely to have been produced by chance and is therefore significant in each case (i.e. the null hypothesis, that there is no correlation, is rejected). However the S97$_{W13}$ sample fails this test. The strongest correlation is between La and Ba, with the La to Eu correlation being of a similar magnitude and both with values of the correlation co-efficient (r) being greater than 0.7, which indicates a strong positive linear relationship. 

For W13 the correlation between Ba and Eu is also significant based on the p-value, but the value of r indicates only a moderate positive linear relationship. The graph itsself shows a great deal of dispersion with at least three extrema, unlike the comparably clean (albeit smaller sample) relation of La to Eu. There is some dispersion in the La and Ba relation. The correlations of La to Eu have been confirmed for three other GCs, as well as M15, in the investigation of literature sources in \citet{Roederer2011}.

However for both samples, there does appear to be evidence of a bimodal distribution in [Ba/H] as seen in Figure~\ref{fig:labaeu}c \& d. For W13, at [Ba/H]$\sim$--2.15 there is a clear separation in the distribution, with the low Ba abundance group being more dispersed in both Ba and Eu than the high Ba abundance group, which appears to have a much tighter distribution (ignoring the high Ba extrema).  For S97$_{W13}$ the separation occurs at [Ba/H]$\sim$--2.30 with the low Ba abundance group showing a more scatter though uniform distribution while the high Ba abundance group has a tight core and is more dispersed in extrema. But the W13 and S97$_{W13}$ samples are in reasonable agreement, if globally offset, in the two groupings so as to emphasise rather than diminish the separation in [Ba/H]. 

In Figure~\ref{fig:labaeu}c the upper limits on the determination of Eu for 6 noisy W13 spectra are shown in red and also fall so as to agree with the bimodal distribution. The upper limits on the two sets of summed spectra are also shown in grey. The two sets of spectra were selected as having components of low Ba abundances in order to explore the low Eu regime, the key line of which was not resolved in the majority of the individual W13 spectra. One set (with higher S/N) clearly falls within the low Ba abundance mode. The second set, comprised of much noisier spectra, lies well below this mode and given how noisy the spectra are should probably be discounted. 

\subsection{Bimodal Ba distribution in M15?}
The existence of a bimodal distribution in Ba and Eu in M15 was raised in \citet{Sneden1997}. A re-analysis of key sets of stars from \citet{Sneden1997}, \citet{Sneden2000a}, \citet{Preston2006} and \citet{Otsuki2006} in the study of \citet{Sobeck2011} determined that such a bimodal distribution is not evident but rather that there was a continuum of increasing Ba with increasing Eu.

\begin{figure*}[!t]
\centering
\begin{minipage}{185mm}  
\hspace{-2cm}
\includegraphics[width=220mm]{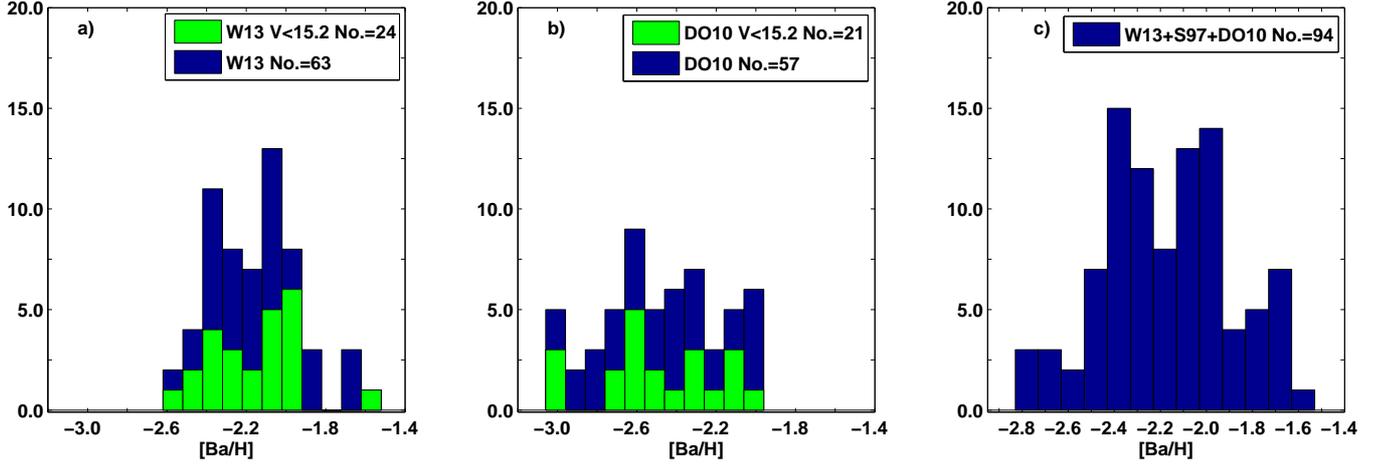}
\caption{Histograms exploring the three key samples of M15 giants in [Ba/H]: a) All 63 W13 stars (blue) and 24 bright W13 stars (green); b) All 57 DO10$_{W13}$ stars (blue) and 21 bright DO10$_{W13}$ stars (green); and c) the W13 sample combined with the further 12 S97$_{W13}$ stars and the further 19 DO10$_{W13}$ stars with offsets applied.} \label{fig:bah_hist}
\end{minipage}
\end{figure*}

To investigate this further Figure~\ref{fig:bah_hist} shows a series of histograms of [Ba/H] for the W13 and DO10$_{W13}$ samples. Figure~\ref{fig:bah_hist}a is the histogram of the W13 sample with a binsize of 0.1~dex (blue). This is the largest M15 sample of giants star to-date that have been analysed for their Ba abundances, which also well samples along the RGB. A bimodal distribution is clearly seen. The histogram of the subset of bright stars from W13 (V $<$ 15.2) is also shown and the bimodal distribution is still clear. These stars are singled out as having the lowest associated error and hence most accurate abundnace determination.

Figure~\ref{fig:bah_hist}c replicates Figure~\ref{fig:bah_hist}a but for the DO10$_{W13}$ sample. This similarly large sample of 57 giant stars also well samples along the giant branch of M15. There is no clear bimodal distribution but also no clear single gaussian distribution either. The distribution appears rectangular in that all the bins, including the extrema, are similarly populated. That the extrema bins are thus populated argues against this distribution being a simple gaussian. Even the bright subsample for DO10$_{W13}$ does not clarify the distribution. It is noted that the range of values is offset to W13 to more Ba-poor values as expected.

Figure~\ref{fig:bah_hist}c combines W13 with the further 12 stars from S97$_{W13}$ and the 19 stars DO10$_{W13}$ by applying the offsets specified above which corrected the S97$_{W13}$ stars and the DO10$_{W13}$ stars to the W13 system. The overall distribution is bimodal but the extrema bins are also well populated. Hence the characteristics of the W13 (bimodal) and DO10 (well-populated extrema bins) are clear and do not erase the effects of each other. However as the crossover samples have a noticable scatter between the Ba results, which implies that per star the Ba abundances from each line and study are not in good agreement, then combining the three does not produce a sample that can be considered to be internally consistent (See Section~\ref{sec:stellarabundanalysis}).

The threshold of the bimodal distribution in the W13 lies at [Ba/H]$\sim$--2.20~dex separating the sample into two modes, Mode I being the low Ba abundance mode and Mode II being the high Ba abundance mode. Table~\ref{tab:modemeans} lists the mean and standard deviation for the three samples of W13, DO10$_{W13}$ and S97$_{W13}$. Statistics are also given for the two modes for W13, DO10$_{W13}$ (assuming a threshold for DO10$_{W13}$ of [Ba/H]$\sim$--2.47~dex) and S97$_{W13}$ (assuming a threshold of [Ba/H]$\sim$--2.36~dex).

\begin{table*}[htbp]
\centering
\caption{Mean and standard deviations of [Fe/H], [Ba/H], [Eu/H] and [La/H] derived (where possible) for the three samples of W13, DO10$_{W13}$ and S97$_{W13}$ for each whole sample and separated into the respective low (I) and high (II) Ba modes. The number of stars for each calculation is also given.}
\begin{tabular}{ccccccccccc}
\hline
 & N & [Fe/H] & N$_{Ba}$ & [Ba/H] & N$_{Eu}$ & [Eu/H] & N$_{La}$ & [La/H] & [Ba/Eu] & [La/Eu] \\ 
\hline\hline
W13 & 63 & --2.34$\pm$0.06 & 63 & --2.20$\pm$0.26 & 20 & --1.70$\pm$0.13 & 13 & --2.04$\pm$0.16 & --0.50$\pm$0.29 & --0.24$\pm$0.21 \\ 
DO10$_{W13}$ & 57 & --2.33$\pm$0.06 & 57 & --2.47$\pm$0.31 & - & -  & -  & -  & -  & -  \\ 
S97$_{W13}$ & 18 & --2.33$\pm$0.03 & 18 & --2.36$\pm$0.28 & 18 & --1.78$\pm$0.17 & -  & -  & --0.58$\pm$0.33 & -  \\ 
 &  &  &  &  &  &  &  &  &  &  \\ 
W13 Mode I & [Ba/H]$<$--2.20 & --2.36$\pm$0.06 & 30 & --2.41$\pm$0.16 & 7 & --1.80$\pm$0.08 & 5 & --2.19$\pm$0.13 & --0.61$\pm$0.18 & --0.39$\pm$0.15 \\ 
W13 Mode II & [Ba/H]$\ge$--2.20 & --2.33$\pm$0.05 & 33 & --2.00$\pm$0.16 & 13 & --1.65$\pm$0.13 & 8 & --1.95$\pm$0.11 & --0.35$\pm$0.21 & --0.30$\pm$0.17 \\ 
 &  &  &  &  &  &  &  &  &  &  \\ 
DO10$_{W13}$ Mode I & [Ba/H]$<$--2.47 & --2.33$\pm$0.05 & 29 & --2.74$\pm$0.18 & - & -  & -  & -  & -  & -  \\ 
DO10$_{W13}$ Mode II & [Ba/H]$\ge$--2.47 & --2.34$\pm$0.07 & 28 & --2.22$\pm$0.16 & - & -  & -  & -  & -  & -  \\ 
 &  &  &  &  &  &  &  &  &  &  \\ 
S97$_{W13}$ Mode I & [Ba/H]$<$--2.36 & --2.34$\pm$0.01 & 10 & --2.56$\pm$0.12 & 10 & --1.92$\pm$0.15 & -  & -  & --0.64$\pm$0.19 & -  \\ 
S97$_{W13}$ Mode II & [Ba/H]$\ge$--2.36 & --2.33$\pm$0.05 & 8 & --2.11$\pm$0.19 & 8 & --1.76$\pm$0.18 & -  & -  & --0.35$\pm$0.26 & -  \\ 
\hline
\end{tabular}
\label{tab:modemeans}
\end{table*}

The [Fe/H] values stay relatively constant across all means and all samples with very low associated spread reflecting the tight spread obtained in C09. While the absolute mean values for Ba are offset from each other between the three studies all show the expected large spread in values. When considering the modes the spread of the [Ba/H] values are on a similar magnitude for each of the samples. 

Regarding Eu, W13 and S97$_{W13}$ show good agreement in the mean for each whole sample and for the mean of Mode I. Mode II, the high Ba mode, shows enhanced Eu for S97$_{W13}$ compared to the mean of Mode II for W13. The spread in Eu values within each mode is of similar magnitude and not significantly smaller than that of the spread in the Ba values. While La abundances are only available for W13 the spread on the whole sample and for each mode are again similar in magnitude to the corresponding spread in Eu.

Certainly for each sample and each element, there is a clear separation in mean values between the modes. Even for Eu and La the separation between the means of Mode I and Mode II are outside the 1~$\sigma$ limit. 

\subsection{{\it s}-process residuals: [Ba/Eu], [La/Eu]}

\begin{figure*}[!t]
\centering
\begin{minipage}{180mm}  
\hspace{-2cm}
\includegraphics[width=220mm]{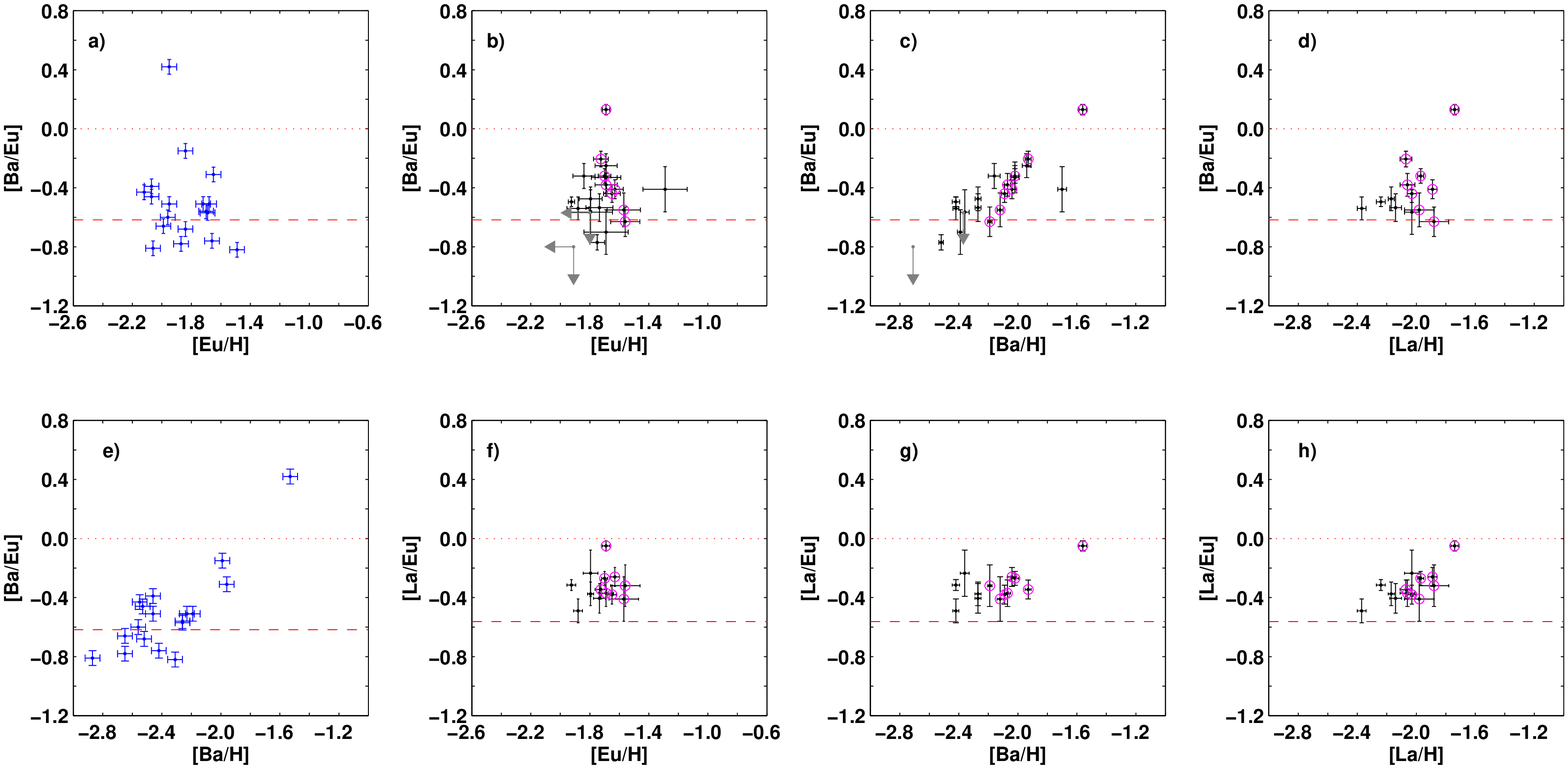}
\caption{a) [Ba/Eu] against [Eu/H] for S97$_{W13}$ (blue).  Dotted red line indicates the solar abundance. Dashed red line indicates the expected abundance from {\it r}-process only yields \citep{Simmerer2004}. b) [Ba/Eu] against [Eu/H] for the W13 sample (black). Grey arrows are the upper limits derived from the set of summed spectra. Magenta circles identify those points with [Ba/H] greater than --2.20~dex and that have measurements for Ba, La and Eu. c) As for b) but for [Ba/Eu] against [Ba/H]. d) as for b) but for [Ba/Eu] against [La/H]. e) As for a) but for [Ba/Eu] against [Ba/H]. f) As for b) but for [La/Eu] against [Eu/H]. g) as for b) but for [La/Eu] against [Ba/H]. h) as for b) but for [La/Eu] against [La/H].}\label{fig:balaeu_ssew}
\end{minipage}
\end{figure*}

The combination of the various studies confirmed the correlation of [Ba/H] with [Eu/H], and [La/H] with [Eu/H] and the analysis here has confirmed a bimodal distribution in Ba and Eu and possibly also in La. To further explore these heavy element abundances in M15 Figure~\ref{fig:balaeu_ssew} compares [Ba/Eu] and [La/Eu] to [Eu/H], [Ba/H] and [La/H] for both the W13 (black) and S97$_{W13}$ (blue) samples.

The ratios of [Ba/Eu] and [La/Eu] can be considered as the contribution of the {\it s}-process to the heavy element abundances once the contribution from the {\it r}-process (Eu) has been removed. For W13 in Figure~\ref{fig:balaeu_ssew}b and f there is no trend of either [Ba/Eu] nor [La/Eu] with [Eu/H]. This is not unexpected, as if all the contribution from the {\it r}-process has been removed then the {\it s}-process contribution would not trend with increasing {\it r}-process abundances. This is borne out by the distribution lying closer to the estimated solar {\it r}-process only yield abundance (indicated as red dashed lines \citep{Simmerer2004}) than to the solar abundance (red dotted lines), which is a combination of both {\it r}- and {\it s}-process contributions. However the distribution does not lie exactly on the estimated solar {\it r}-process only yield implying that there is some contribution of the heavy elements from some other process. This is replicated in Figure~\ref{fig:balaeu_ssew}a for 
the S97$_{W13}$ sample.

Figure~\ref{fig:balaeu_ssew}c is particularly interesting as it shows whether [Ba/Eu] shows any trend with increasing [Ba/H], thus s-process contribution increasing with s-process abundance. The bimodal distribution in Ba can be seen as a plateau for Mode I, implying that this mode has had no further heavy element contribution from the {\it s}-process, but a distinct linear trend of increasing [Ba/Eu] with [Ba/H] for Mode~II. Both of these distributions for each mode are replicated in the S97$_{W13}$, with the linear trend in Mode II appearing to extend to the outliers of each sample. This would imply that the stars in Mode~I have undergone further Ba enhancement processes (i.e. the {\it s}-process).

Is this linear relation in Mode II real? As La and Ba are both {\it s}-process elements then an interchange of these parameters should show the same linear relation. However, as La could only be measured for 13 of the W13 sample and was not available as an equivalent width measurement in S97, then the sample for comparison is significantly reduced. In order to aid the investigation the 8 stars with all three Ba, La and Eu abundances that lie within Mode~II are identified in all the W13 graphs of Figure~\ref{fig:balaeu_ssew} as magenta circles. For Figure~\ref{fig:balaeu_ssew}d there is no convincing linear trend, despite the one outlier. For Figure~\ref{fig:balaeu_ssew}g and h, if one excludes the high La outlier, then there is no strong argument for a linear relation. 

An examination of the parameters of the stars within Mode~II showed no obvious trend of Ba abundance with $\xi$. However for five of these objects there is a distinct linear trend with $T_{\textrm{eff}}$ and $\log g$. Given the almost linear relations between V and these two parameters, and the linear relation developed here between $\xi$ and $\log g$ it is likely that these are the cause of the observed linear trend between [Ba/Eu] with [Ba/H]. The error analysis shows the sensitivity of the analysed Ba line to these parameters, which is accentuated for stronger/more enhanced features. Hence this is the most likely reason why this relation is observed in the high Ba Mode~II, not the low Ba Mode~I. 

The La line is weaker and hence not so sensitive to $\xi$. While there are fewer points, the lack of a linear trend should be taken as confirmation that the linear trend in Ba is not real. However, the location of the [La/Eu] distribution with respect to the {\it r}-process only yields and the solar abundance does not exclude some contribution from the {\it s}-process or some other process that produces heavy elements.

\section{Heavy and light element contributions within M15}\label{sec:heavylightdis}
The existence of two modes in Ba and Eu abundances in M15 was first proposed in S97. For both modes the mean [Ba/Eu] determined in S97 were found to be almost exactly the same ($<$[Ba/Eu]$>\sim$--0.40$\pm$0.05~dex) and indicative of a pollution history dominated by contributions from the {\it r}-process. In this study there is a separation in the mean [Ba/Eu] values of $\sim$0.3~dex outside the uncertainties (see Table~\ref{tab:modemeans}), with the value for Mode~I being less than that for Mode~II. In the S97$_{W13}$ sample similar values to W13 were found and so a similar difference outside the uncertainties. But both indicate no extra contribution from the s-process, therefore this study agrees with S97 that both modes have an enrichment history dominated by the {\it r}-process but that Mode~I has been less enriched in the heavy elements than Mode~II.  

\begin{figure}[!h]
\centering
\begin{minipage}{90mm}  
\hspace{-0.3cm}
\includegraphics[width=95mm]{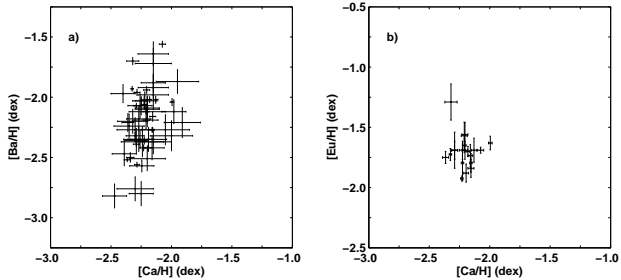}
\caption{Comparison of light with heavy element abundances for the W13 sample: a) [Ba/H] against [Ca/H]; b) [Eu/H] against [Ca/H].}\label{fig:baeu_ca}
\end{minipage}
\end{figure}

However when considering the s-process residuals as represented by [La/Eu] both modes have values in close agreement and well within the uncertainties, with Mode~I being slightly less enhanced. Given the greater consistency that can be obtained from La measurements this can be well relied upon, but still supports the above conclusion that both Modes are r-process dominated with a possible greater enrichment in Mode~II.

S00b confirmed the conclusions of S97 on a larger sample of M15 stars for which Ba only was determined of the heavy elements. S00a investigated in detail several heavy element abundances for three of the S97 giants, two that lie in the proposed Mode~I and one in Mode~II. The signatures that they found were confirmation of {\it r}-process dominated abundances.

\begin{figure*}[!t]
\centering
\begin{minipage}{180mm}  
\hspace{-2cm}
\includegraphics[width=220mm]{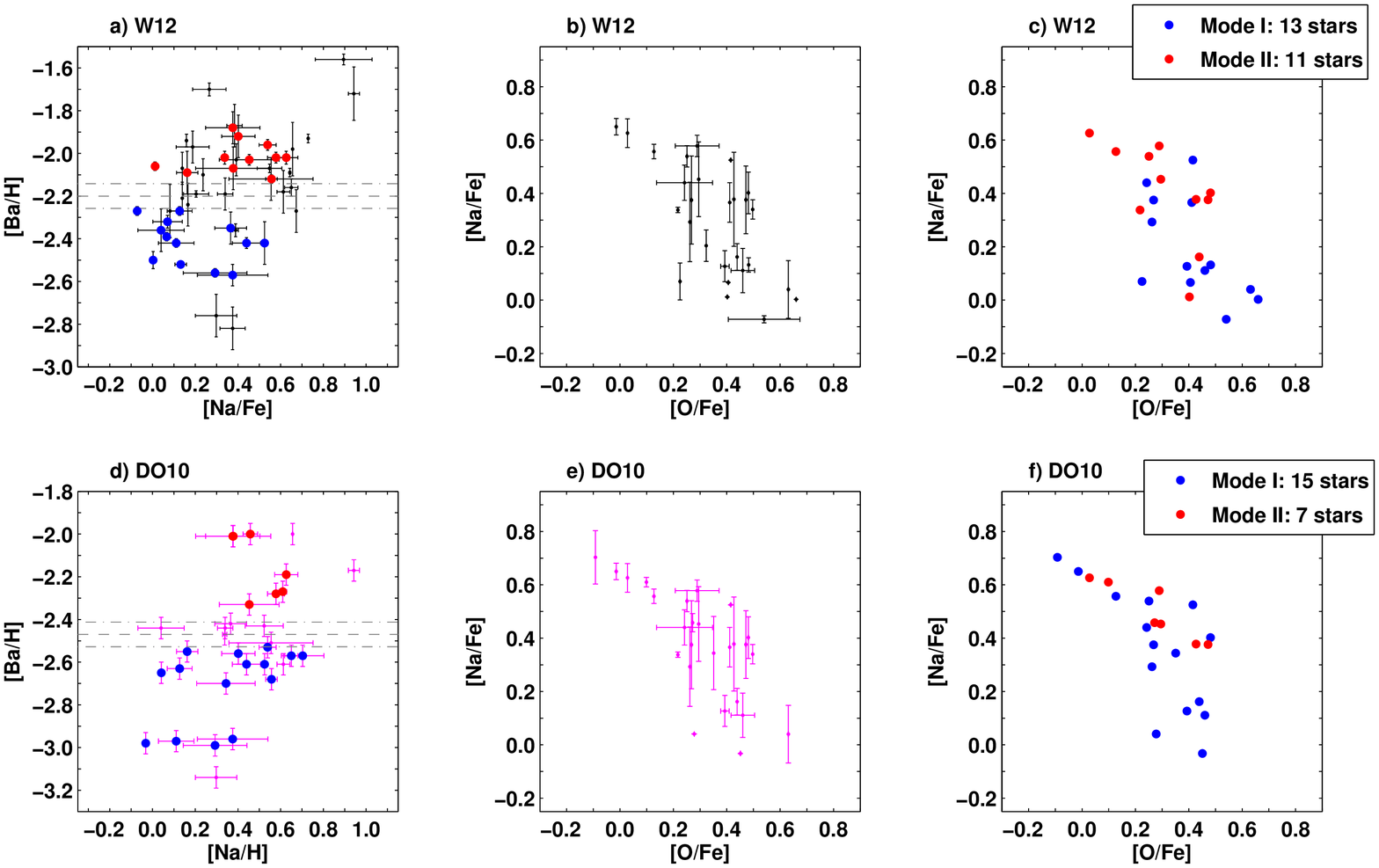}
\caption{a) [Ba/Eu] against [Na/Fe] for 49 stars in the W13 sample (black).  Dashed grey line indicates the W13 mode threshold. Dash-dotted grey lines indicates the mean $\sigma$[Ba/H] above and below the threshold. Blue points identify those clearly in Mode I, red points are those clearly in Mode II. b) [Na/Fe] against [O/Fe] for the W13 stars with both measurements. c) As for b) but the red and blue points from a). d) to f) are as for a) to c) but for  DO10$_{W13}$ (magenta).}\label{fig:ba_na}
\end{minipage}
\end{figure*}

The heavy elements can be paired with light element counterparts based on their common sites of nucleosynthesis in order to constrain the exact nature of the source of these chemical abundance patterns. In particular, the {\it r}-process and $\alpha$-capture processes (producing Ca) can both occur in Supernova Type~II. Similarly proton-capture processes (producing Na) occur during the evolution up the AGB, as does the {\it s}-process in the thermally pulsing stage.

S97 determined that neither the Ba nor Eu abundances correlated with the $\alpha$-element abundances. In this study no correlation was found between Ba and Ca, nor between Eu and Ca supporting the result found in S97 as shown in Figure~\ref{fig:baeu_ca}a and b. The conclusion drawn in S97, that is supported here, was that the mass of the star resulting in the Type II Supernova that produced the {\it r}-process element was insufficient to simultaneously produce the $\alpha$-elements. The Ba and Eu abundances derived here fall within the range of values of collated data and model yields in \citet{Chen2006}. Based on this the two modes of M15 are both indicative of pollution by Type II supernova. As concluded in \citet{Chen2006} and references therein, Supernova Type II associated with stars of progenitor mass between 20~M$_{\odot}$ and 40~M$_{\odot}$ are the main source of r-process elements in the early galaxy.

With regards to the {\it s}-process and proton-capture elements, Na was measured for S97, S00b, O06 and Sk11 but no correlation between Na and either Ba or La was found. Similarly no correlation was found between the {\it s}-process elements measured here and the corresponding Na abundance measured in C09. This is seen in Figure~\ref{fig:ba_na}a and c where both Na and Ba are highly dispersed for both the W13 and DO10$_{W13}$ samples.

Because the Na and Ba do not correlate then these elements were produced by different sources which polluted the medium from which the observed stars formed, as neither Na nor Ba can be produced in these observed stars themselves. Hence this limits the potential AGB polluter to those with sufficient mass to produce Na but not Ba (in the thermally pulsing stage) and is evidence for at least two generations of stars in M15 \citep{Carretta2009a}. This is borne out by comparison to theoretical yields of s-process element abundances from AGB models. \citet{Lugaro2012} present s-process yields calculated from AGB models at [Fe/H]=--2.3~dex, comparable to the metallicity of M15. The expected [Ba/Eu] abundances for a range of stellar masses all exceed [Ba/Eu]=0.0~dex (see tables 3 and 4 of that paper). This is therefore in support of the conclusion here that any pollution from AGB in M15 contributed only to the light element abundances, not to the s-process element abundances.

The Na-O anticorrelation is a well-studied characteristic of GCs \citep[][and references therein]{Carretta2010b}. It is proposed that the first generation of stars inherited the signature of the supernova nucleosynthesis whereby the O abundances are enhanced to above solar values while the Na is depleted below solar values. During the lifetimes of the massive stars of this population the material is processed in hot H-burning such that the O is depleted and the Na is enhanced. This level of depletion and enhancement depends on the characteristics of the individual stars. Hence a continuum is seen for the Na-O anticorrelation as different levels of contribution are made to the second generation star-forming material depending on the contributing first generation star. Therefore the second generation of stars bear the signature of the nucleosynthetic processing that occurred within the first generation stars. 

Figure~\ref{fig:ba_na}b and e show the Na-O distribution for the W13 and DO10$_{W13}$ samples respectively. The Na-O anticorrelation is clear in both samples. Hence evidence of at least two generations of stars in M15. So where does the Ba bimodal distribution fit in? Does one of the modes represent the first generation and the other the second generation? If so then one mode would show the signature of super-solar O and sub-solar Na abundances, while the other is depleted in O and enhanced in Na.

Figure~\ref{fig:ba_na}c and d decompose the Na-O distribution into the two Ba modes for W13 and DO10$_{W13}$ respectively, as defined in Figure~\ref{fig:ba_na}a and c. In both instances we restricted the modes to those stars most likely to be in each mode, and so did not include those stars near the mean of each sample as illustrated by the red dashed lines in the figures. What is most clearly seen in the W13 sample is that both Ba modes show the anti-correlation in Na-O. In the DO10$_{W13}$ sample the anticorrelation is clear for Mode I but more truncated in Mode II. The two modes are less evenly populated for DO10$_{W13}$, as was seen in Figure~\ref{fig:bah_hist} and so the sample size for Mode II may be insufficient.

Based on the W13 sample, the polluters of Mode II appear to have undergone more processing of the material because the correlation extends further into the enhanced Na,depleted O regime.

Taking the two Ba modes and then the two stellar generations within each mode there is a potential scenario for the formation of M15 which involves four populations of stars:
\begin{itemize}
 \item Mode I: Primordial showing SN enrichment (relatively poor Ba \& Eu, sub-solar Na, super-solar O)
 \item Mode I: Second generation showing SN enrichment and H-burning products (relatively poor Ba \& Eu, enhanced Na, depleted O)
 \item Mode II: Primordial showing SN enrichment (relatively rich Ba \& Eu, sub-solar Na, super-solar O)
 \item Mode II: Second generation showing SN enrichment and H-burning products (relatively rich Ba \& Eu, enhanced Na, depleted O)
\end{itemize}

In such a scenario each mode was initially polluted by the explosive nucleosynthesis of massive stars with slightly different heavy element yields resulting in the bimodal distribution that has been observed in the W13 sample. This difference in composition may have had the follow on effect of greater processing of the O to Na for the higher Ba mode (II). Integral to formulating such a scenario are the timescales that have been involved. Did these two modes evolve simultaneously, and somehow disconnectedly either separately or within the same cluster space, or did one population evolve followed by the other?

Alternatively these signatures could reflect localised events. The stars all formed on the same timescale but the initial primordial material was not well-mixed such that stars in one region were more enhanced (depleted) in the heavy elements depending on the level of polllution by the SN. If the bimodal distribution is real then it would appear that the SN events were restricted to two types. If the bimodal distribution is not real, and there is a continuum of heavy element abundances, as might be argued based on the DO10$_{W13}$ sample, then this restriction is relaxed and the Na-O anticorrelation may be representative of a single mode scenario involving a single primordial population and a single second generation population.

Sk11 investigated the heavy element abundances for 7 RHB and 3 RGB stars in M15, in combination with a re-analysis of the S97 stars using the improved analysis techniques, and found no evidence of a bimodal distribution in Ba and Eu for the sample. Detailed investigation was given into light as well as heavy {\it s}-process elements whereby the signatures of the 10 key stars showed evidence of enrichment in the heavy elements that was not attributable to either the {\it r}- nor {\it s}-process but potentially due to a Lighter Element Primary Process (LEPP). Due to the wavelength range of the observations analysed in the current paper there were no spectral features of any light {\it s}-process elements (Sr, Y, Zr) present, hence this line of research could not be pursued. 

The above investigation of the s-process residual with [Ba/H], [La/H] and [Eu/H] confirmed that both modes are dominated by contributions from the r-process, hence pollution by explosive nucleosynthesis in massive stars. Some small contribution to the heavy element abundances by another process cannot be discounted as the residuals for both modes lie above the r-process only yield limit, although predictions of heavy element yields from explosive nucleosynthesis have their own set of uncertainties. As such the distribution of heavy elements in this sample of M15 giant stars implies one mode that is metal poor with low Ba, La and Eu abundances, and a second mode which is slightly less metal poor ($\Delta$[Fe/H]=+0.03~dex) with relatively higher Ba, La and Eu abundances that all have similar dispersions on the order of the dispersion for Fe. 

In effect the analysis carried out here has drawn the same conclusions as S97, S00a and S00b that there is a bimodal distribution in the Ba, Eu, and now potentially La, abundances in M15 giant stars and that both modes therefore have a heavy element enrichment history that is dominated by the {\it r}-process. The addition of the Na-O abundances from C09 have added another layer of information which potentially describes a very complex formation scenario for this very interesting GC.

Throughout Figures~\ref{fig:labaeu} and \ref{fig:balaeu_ssew} there is an outlier (C09 ID 42262) in the high Ba Mode that is enhanced in Ba and La to a level which is significantly greater than the rest of the W13 stars. However, the Eu abundance is fairly typical of that mode. The s-process residuals for this star place it close to the solar value. Given the conclusions above and that because this is an RGB star it cannot be producing the heavy elements internally, the most likely scenario is that this star has been the recipient of mass transfer from a companion star that has resulted in the observed enhancments. The fact that the star is not further enhanced in Eu (the r-process element) and both Ba and La are heavy s-process elements implies that the polluting star had evolved through the thermally pulsing AGB stage. 

One other star (C09 ID 37931) in the W13 sample is similarly enhanced in Ba but no La abundance could be measured. This star is also enhanced in Eu, albeit with large uncertainties. There are potentially two such stars within the S97 sample as well.

\section{Conclusions}\label{sec:conclusion}
This is the largest study of barium in giant stars in M15 to-date. The sample, based on the original programme in C09, is well-sampled along the RGB. La and Eu abundances were determined for a subset of this sample furthering this investigation into heavy element abundances in M15. In parallel the equivalent width measurements from DO10 and S97 were re-analysed on the C09 scale which provided an interesting comparison for key element distributions.

The stellar parameters for these objects were take from C09, except for $\xi$ which was re-determined based on a $\xi$--$\log g$ relation derived from a subsample of stars for which a reasonable sample of Fe I lines could be measured. The key Ba line under analysis was shown to be highly sensitive to $\xi$ and so a great effort was made to reduce the associated errors.

A bimodal distribution of Ba, Eu, and potentially La, was found in the W13 sample in agreement with previous studies. Based on the W13 results, an examination of the {\it r}- and {\it s}-process contributions to these heavy element abundances leads to the two modes having similar pollution scenarios of a single burst of explosive nucleosynthese by massive stars (for example, Supernova Type II) that enriched the interstellar medium from which the stars in each mode formed. The enrichment was to different degrees between the modes.

Further enrichment by another process that produces heavy elements can not be discounted but the wavelength region observed here did not allow for analysis of other heavy element features.

For both modes a Na-O anticorrelation was observed which is indicative of two stellar generations in each mode: a primodial and second generation in Mode I, and a primordial and second generation in Mode II. A scenario that can account for such a set of stellar populations is necessarily complex. If the Ba distribution is more indicative of a continuum of Ba and Eu abundances then a single mode of stellar evolution could explain the observed Na-O anticorrelation.

While this study strongly supports the argument for a bimodal distribution in Ba, La and Eu for M15, there are inherent uncertainties in the measurement of the key Ba line due to its sensitivity to $\xi$. La would be a better candidate as it has no such sensitivites. Hence re-observing the entire sample considered here at higher resolution and S/N in order to complete the La and Eu abundances for the sample would prove definitive. Extending the observed wavelength regions to include features of the light {\it s}-process peak (Y, Sr, Zr) would enrich the analysis as to the potential contribution from other heavy element processes.

\begin{acknowledgements}
This research has made use of the SIMBAD database, operated at CDS, Strasbourg, France.
\end{acknowledgements}

\bibliographystyle{aa}


\begin{thebibliography}{30}
\expandafter\ifx\csname natexlab\endcsname\relax\def\natexlab#1{#1}\fi

\bibitem[{{Alonso} {et~al.}(1999){Alonso}, {Arribas}, \&
  {Mart{\'{\i}}nez-Roger}}]{Alonso1999}
{Alonso}, A., {Arribas}, S., \& {Mart{\'{\i}}nez-Roger}, C. 1999, A$\&$AS, 140,
  261

\bibitem[{{Andrievsky} {et~al.}(2009){Andrievsky}, {Spite}, {Korotin}, {Spite},
  {Fran{\c c}ois}, {Bonifacio}, {Cayrel}, \& {Hill}}]{Andrievsky2009}
{Andrievsky}, S.~M., {Spite}, M., {Korotin}, S.~A., {et~al.} 2009, \aap, 494,
  1083

\bibitem[{{Busso} {et~al.}(2001){Busso}, {Gallino}, {Lambert}, {Travaglio}, \&
  {Smith}}]{Busso2001}
{Busso}, M., {Gallino}, R., {Lambert}, D.~L., {Travaglio}, C., \& {Smith},
  V.~V. 2001, ApJ, 557, 802

\bibitem[{{Carretta} {et~al.}(2009{\natexlab{a}}){Carretta}, {Bragaglia},
  {Gratton}, \& {Lucatello}}]{Carretta2009b}
{Carretta}, E., {Bragaglia}, A., {Gratton}, R., \& {Lucatello}, S.
  2009{\natexlab{a}}, \aap, 505, 139

\bibitem[{{Carretta} {et~al.}(2009{\natexlab{b}}){Carretta}, {Bragaglia},
  {Gratton}, {Lucatello}, {Catanzaro}, {Leone}, {Bellazzini}, {Claudi},
  {D'Orazi}, {Momany}, {Ortolani}, {Pancino}, {Piotto}, {Recio-Blanco}, \&
  {Sabbi}}]{Carretta2009a}
{Carretta}, E., {Bragaglia}, A., {Gratton}, R.~G., {et~al.} 2009{\natexlab{b}},
  \aap, 505, 117

\bibitem[{{Carretta} {et~al.}(2010){Carretta}, {Bragaglia}, {Gratton},
  {Recio-Blanco}, {Lucatello}, {D'Orazi}, \& {Cassisi}}]{Carretta2010b}
{Carretta}, E., {Bragaglia}, A., {Gratton}, R.~G., {et~al.} 2010, \aap, 516,
  A55

\bibitem[{{Chen} {et~al.}(2006){Chen}, {Zhang}, {Chen}, {Cui}, \&
  {Zhang}}]{Chen2006}
{Chen}, Z., {Zhang}, J., {Chen}, Y., {Cui}, W., \& {Zhang}, B. 2006, \apss,
  306, 33

\bibitem[{{Davidson} {et~al.}(1992){Davidson}, {Snoek}, {Volten}, \&
  {Doenszelmann}}]{Davidson1992}
{Davidson}, M.~D., {Snoek}, L.~C., {Volten}, H., \& {Doenszelmann}, A. 1992,
  \aap, 255, 457

\bibitem[{{D'Orazi} {et~al.}(2010){D'Orazi}, {Gratton}, {Lucatello},
  {Carretta}, {Bragaglia}, \& {Marino}}]{DOrazi2010}
{D'Orazi}, V., {Gratton}, R., {Lucatello}, S., {et~al.} 2010, \apjl, 719, L213

\bibitem[{{Edvardsson} {et~al.}(1993){Edvardsson}, {Andersen}, {Gustafsson},
  {Lambert}, {Nissen}, \& {Tomkin}}]{Edvardsson1993}
{Edvardsson}, B., {Andersen}, J., {Gustafsson}, B., {et~al.} 1993, \aap, 275,
  101

\bibitem[{{Gratton} {et~al.}(2012){Gratton}, {Carretta}, \&
  {Bragaglia}}]{Gratton2012}
{Gratton}, R.~G., {Carretta}, E., \& {Bragaglia}, A. 2012, \aapr, 20, 50

\bibitem[{{Iliadis}(2007)}]{Iliadis07book}
{Iliadis}, C. 2007, Nuclear Physics of Stars (Wiley-VCH)

\bibitem[{{Kirby} {et~al.}(2009){Kirby}, {Guhathakurta}, {Bolte}, {Sneden}, \&
  {Geha}}]{Kirby2009}
{Kirby}, E.~N., {Guhathakurta}, P., {Bolte}, M., {Sneden}, C., \& {Geha}, M.~C.
  2009, \apj, 705, 328

\bibitem[{{Lawler} {et~al.}(2001{\natexlab{a}}){Lawler}, {Bonvallet}, \&
  {Sneden}}]{Lawler2001a}
{Lawler}, J.~E., {Bonvallet}, G., \& {Sneden}, C. 2001{\natexlab{a}}, \apj,
  556, 452

\bibitem[{{Lawler} {et~al.}(2001{\natexlab{b}}){Lawler}, {Wickliffe}, {den
  Hartog}, \& {Sneden}}]{Lawler2001b}
{Lawler}, J.~E., {Wickliffe}, M.~E., {den Hartog}, E.~A., \& {Sneden}, C.
  2001{\natexlab{b}}, \apj, 563, 1075

\bibitem[{{Lodders}(2003)}]{Lodders2003}
{Lodders}, K. 2003, \apj, 591, 1220

\bibitem[{{Lugaro} {et~al.}(2012){Lugaro}, {Karakas}, {Stancliffe}, \&
  {Rijs}}]{Lugaro2012}
{Lugaro}, M., {Karakas}, A.~I., {Stancliffe}, R.~J., \& {Rijs}, C. 2012, \apj,
  747, 2

\bibitem[{{Otsuki} {et~al.}(2006){Otsuki}, {Honda}, {Aoki}, {Kajino}, \&
  {Mathews}}]{Otsuki2006}
{Otsuki}, K., {Honda}, S., {Aoki}, W., {Kajino}, T., \& {Mathews}, G.~J. 2006,
  \apjl, 641, L117

\bibitem[{{Preston} {et~al.}(2006){Preston}, {Sneden}, {Thompson}, {Shectman},
  \& {Burley}}]{Preston2006}
{Preston}, G.~W., {Sneden}, C., {Thompson}, I.~B., {Shectman}, S.~A., \&
  {Burley}, G.~S. 2006, \aj, 132, 85

\bibitem[{{Roederer}(2011)}]{Roederer2011}
{Roederer}, I.~U. 2011, \apjl, 732, L17

\bibitem[{{Rutten}(1978)}]{Rutten1978}
{Rutten}, R.~J. 1978, \solphys, 56, 237

\bibitem[{{Short} \& {Hauschildt}(2006)}]{Short2006}
{Short}, C.~I. \& {Hauschildt}, P.~H. 2006, \apj, 641, 494

\bibitem[{{Simmerer} {et~al.}(2004){Simmerer}, {Sneden}, {Cowan}, {Collier},
  {Woolf}, \& {Lawler}}]{Simmerer2004}
{Simmerer}, J., {Sneden}, C., {Cowan}, J.~J., {et~al.} 2004, \apj, 617, 1091

\bibitem[{{Sneden}(1973)}]{MOOG}
{Sneden}, C. 1973, PhD thesis, University of Texas at Austin

\bibitem[{{Sneden} {et~al.}(2008){Sneden}, {Cowan}, \& {Gallino}}]{Sneden2008}
{Sneden}, C., {Cowan}, J.~J., \& {Gallino}, R. 2008, \araa, 46, 241

\bibitem[{{Sneden} {et~al.}(2000{\natexlab{a}}){Sneden}, {Johnson}, {Kraft},
  {Smith}, {Cowan}, \& {Bolte}}]{Sneden2000a}
{Sneden}, C., {Johnson}, J., {Kraft}, R.~P., {et~al.} 2000{\natexlab{a}},
  \apjl, 536, L85

\bibitem[{{Sneden} {et~al.}(1997){Sneden}, {Kraft}, {Shetrone}, {Smith},
  {Langer}, \& {Prosser}}]{Sneden1997}
{Sneden}, C., {Kraft}, R.~P., {Shetrone}, M.~D., {et~al.} 1997, \aj, 114, 1964

\bibitem[{{Sneden} {et~al.}(2000{\natexlab{b}}){Sneden}, {Pilachowski}, \&
  {Kraft}}]{Sneden2000b}
{Sneden}, C., {Pilachowski}, C.~A., \& {Kraft}, R.~P. 2000{\natexlab{b}}, \aj,
  120, 1351

\bibitem[{{Sobeck} {et~al.}(2011){Sobeck}, {Kraft}, {Sneden}, {Preston},
  {Cowan}, {Smith}, {Thompson}, {Shectman}, \& {Burley}}]{Sobeck2011}
{Sobeck}, J.~S., {Kraft}, R.~P., {Sneden}, C., {et~al.} 2011, \aj, 141, 175

\bibitem[{{Worley} {et~al.}(2012){Worley}, {de Laverny}, {Recio-Blanco},
  {Hill}, {Bijaoui}, \& {Ordenovic}}]{Worley2012}
{Worley}, C.~C., {de Laverny}, P., {Recio-Blanco}, A., {et~al.} 2012, \aap,
  542, A48

\end{thebibliography}

\onecolumn
\begin{landscape}
\begin{center}
\newdimen\LTcapwidth \LTcapwidth=210mm  
\renewcommand{\arraystretch}{0.75}
\begin{longtable}{cccccccccccccccccc}
\caption{Chemical abundances, uncertainties and number of lines measured here for the sample of 63 M15 giants stars. The C09 stellar parameters, $\xi_{W13}$ and measured S/N are also listed.}\label{tab:chemabund} \\

\hline\hline \\
ID & S/N & $T_{\textrm{eff}}$ & $\log g$ & [Fe/H] & $\xi_{W13}$ & [Ca/Fe] & $\sigma$ & N & [Ni/Fe] & $\sigma$ & [Ba/Fe] & $\sigma$ & [La/Fe] & $\sigma$ & [Eu/Fe] & $\sigma$ & N \\ 
 &  & (K) & (dex) & (dex) & (kms$^{-1}$) & (dex) & (dex) &  & (dex) & (dex) & (dex) & (dex) & (dex) & (dex) & (dex) & (dex) &  \\ 
\hline\\
\endfirsthead

\hline\hline \\
ID & S/N & $T_{\textrm{eff}}$ & $\log g$ & [Fe/H] & $\xi_{W13}$ & [Ca/Fe] & $\sigma$ & N & [Ni/Fe] & $\sigma$ & [Ba/Fe] & $\sigma$ & [La/Fe] & $\sigma$ & [Eu/Fe] & $\sigma$ & N \\ 
 &  & (K) & (dex) & (dex) & (kms$^{-1}$) & (dex) & (dex) &  & (dex) & (dex) & (dex) & (dex) & (dex) & (dex) & (dex) & (dex) &  \\ 
\hline \\
\endhead

\\
\hline \\
\multicolumn{8}{r}{continued on next page} \\ 
\endfoot

\endlastfoot

40825 & 135 & 4313 & 0.65 & -2.33 & 2.25 & 0.10 & 0.02 & 2 & -0.05 & 0.01 & -0.09 & 0.03 & 0.09 & 0.03 & 0.41 & 0.03 & 2 \\ 
4099 & 128 & 4324 & 0.69 & -2.32 & 2.30 & 0.16 & 0.02 & 2 & -0.02 & 0.01 & 0.05 & 0.02 & 0.18 & 0.04 & 0.59 & 0.09 & 2 \\ 
43788 & 118 & 4325 & 0.52 & -2.36 & 2.25 & 0.21 & 0.05 & 2 & 0.16 & 0.01 & 0.09 & 0.02 & 0.19 & 0.02 & 0.57 & 0.08 & 2 \\ 
31914 & 112 & 4338 & 0.58 & -2.20 & 2.10 & 0.21 & 0.02 & 2 & 0.04 & 0.02 & 0.16 & 0.03 & 0.31 & 0.03 & 0.57 & 0.06 & 2 \\ 
41287 & 162 & 4373 & 0.78 & -2.36 & 2.10 & 0.14 & 0.06 & 2 & 0.01 & 0.01 & 0.00 & 0.03 & 0.33 & 0.05 & 0.57 & 0.16 & 2 \\ 
34519 & 161 & 4416 & 0.87 & -2.38 & 2.00 & 0.18 & 0.08 & 2 & 0.06 & 0.01 & -0.04 & 0.02 & 0.00 & 0.03 & 0.50 & 0.08 & 1 \\ 
37215 & 164 & 4445 & 0.98 & -2.31 & 2.1 & -0.02 & 0.04 & 2 & -0.27 & 0.02 & 0.38 & 0.02 & 0.24 & 0.04 & 0.59 & 0.06 & 2 \\ 
34995 & 141 & 4468 & 1.00 & -2.34 & 2.10 & 0.13 & 0.06 & 2 & -0.01 & 0.03 & 0.25 & 0.02 & 0.31 & 0.03 & 0.69 & 0.07 & 2 \\ 
3137 & 152 & 4486 & 1.08 & -2.35 & 2.00 & 0.17 & 0.04 & 2 & 0.03 & 0.02 & 0.33 & 0.03 & 0.38 & 0.02 & 0.65 & 0.05 & 2 \\ 
42262 & 103 & 4528 & 1.13 & -2.33 & 2.0 & 0.26 & 0.03 & 2 & 0.11 & 0.02 & 0.77 & 0.03 & 0.59 & 0.03 & 0.64 & 0.03 & 1 \\ 
26751 & 167 & 4533 & 1.18 & -2.44 & 2.0 & 0.15 & 0.04 & 2 & -0.05 & 0.02 & 0.37 & 0.02 & 0.38 & 0.05 & 0.75 & 0.08 & 1 \\ 
38678 & 143 & 4554 & 1.23 & -2.35 & 2.00 & 0.14 & 0.03 & 2 & -0.01 & 0.02 & 0.23 & 0.02 & 0.37 & 0.10 & 0.78 & 0.11 & 2 \\ 
2792 & 147 & 4567 & 1.26 & -2.32 & 2.20 & 0.12 & 0.04 & 2 & -0.04 & 0.03 & 0.13 & 0.02 & 0.44 & 0.10 & 0.76 & 0.11 & 2 \\ 
39752 & 171 & 4598 & 1.31 & -2.36 & 2.0 & 0.07 & 0.05 & 2 & -0.08 & 0.02 & -0.20 & 0.02 & - & - & 0.30 & 0.08 & 1 \\ 
34456 & 140 & 4621 & 1.37 & -2.30 & 2.0 & 0.17 & 0.03 & 2 & -0.02 & 0.02 & 0.28 & 0.03 & - & - & 0.61 & 0.10 & 1 \\ 
34961 & 156 & 4627 & 1.41 & -2.32 & 2.00 & 0.17 & 0.03 & 2 & -0.02 & 0.03 & 0.16 & 0.04 & - & - & 0.48 & 0.08 & 1 \\ 
39787 & 165 & 4630 & 1.35 & -2.43 & 2.0 & 0.06 & 0.06 & 2 & -0.04 & 0.03 & -0.09 & 0.02 & - & - & 0.68 & 0.05 & 1 \\ 
38329 & 166 & 4664 & 1.44 & -2.37 & 1.80 & 0.17 & 0.03 & 2 & 0.03 & 0.04 & 0.43 & 0.03 & - & - & 0.68 & 0.08 & 1 \\ 
45062 & 146 & 4700 & 1.54 & -2.31 & 1.9 & 0.09 & 0.03 & 2 & -0.08 & 0.04 & 0.29 & 0.03 & - & - & 0.62 & 0.08 & 1 \\ 
20498 & 159 & 4717 & 1.60 & -2.49 & 1.9 & 0.21 & 0.04 & 2 & -0.16 & 0.03 & 0.53 & 0.03 & - & - & - & - & - \\ 
21948 & 167 & 4746 & 1.66 & -2.49 & 1.9 & 0.20 & 0.10 & 2 & -0.07 & 0.03 & 0.10 & 0.02 & - & - & 0.80 & 0.15 & 1 \\ 
28510 & 145 & 4754 & 1.64 & -2.42 & 1.9 & 0.12 & 0.08 & 2 & -0.02 & 0.03 & 0.10 & 0.03 & - & - & - & - & - \\ 
29264 & 146 & 4810 & 1.77 & -2.35 & 1.8 & 0.13 & 0.03 & 2 & -0.06 & 0.03 & 0.29 & 0.02 & - & - & - & - & - \\ 
18815 & 143 & 4832 & 1.83 & -2.29 & 1.8 & -0.05 & 0.09 & 2 & -0.23 & 0.10 & -0.21 & 0.04 & - & - & 0.30 & 0.08 & 1 \\ 
42362 & 129 & 4952 & 2.30 & -2.41 & 1.7 & 0.17 & 0.23 & 2 & 0.05 & 0.10 & -0.16 & 0.05 & - & - & 0.72 & 0.10 & 1 \\ 
29480 & 112 & 4968 & 2.10 & -2.33 & 1.7 & 0.08 & 0.05 & 1 & -0.12 & 0.10 & 0.30 & 0.03 & - & - & - & - & - \\ 
37931 & 113 & 4976 & 2.10 & -2.37 & 1.7 & 0.05 & 0.05 & 1 & -0.09 & 0.10 & 0.67 & 0.03 & - & - & 1.13 & 0.15 & 1 \\ 
31791 & 112 & 4978 & 2.13 & -2.28 & 1.7 & 0.03 & 0.05 & 1 & -0.02 & 0.08 & 0.21 & 0.08 & - & - & - & - & - \\ 
33484 & 103 & 5036 & 2.22 & -2.44 & 1.7 & 0.08 & 0.05 & 1 & -0.15 & 0.08 & 0.23 & 0.08 & - & - & - & - & - \\ 
10329 & 91 & 5060 & 2.25 & -2.38 & 1.7 & 0.22 & 0.10 & 1 & -0.01 & 0.15 & -0.04 & 0.10 & - & - & - & - & - \\ 
27889 & 94 & 5063 & 2.28 & -2.31 & 1.7 & 0.02 & 0.10 & 1 & - & - & -0.11 & 0.10 & - & - & - & - & - \\ 
39741 & 97 & 5076 & 2.29 & -2.42 & 1.7 & 0.12 & 0.15 & 1 & 0.08 & 0.10 & 0.15 & 0.10 & - & - & - & - & - \\ 
42674 & 96 & 5076 & 2.31 & -2.25 & 1.7 & -0.01 & 0.10 & 1 & 0.01 & 0.10 & 0.18 & 0.10 & - & - & - & - & - \\ 
17458 & 86 & 5081 & 2.33 & -2.31 & 1.7 & 0.11 & 0.10 & 1 & -0.03 & 0.10 & 0.22 & 0.10 & - & - & - & - & - \\ 
40086 & 93 & 5087 & 2.36 & -2.32 & 1.7 & 0.06 & 0.10 & 1 & -0.12 & 0.10 & -0.03 & 0.08 & - & - & - & - & - \\ 
36274 & 98 & 5105 & 2.35 & -2.35 & 1.7 & 0.14 & 0.10 & 1 & - & - & 0.16 & 0.08 & - & - & - & - & - \\ 
26759 & 99 & 5106 & 2.34 & -2.35 & 1.7 & 0.15 & 0.10 & 1 & - & - & 0.25 & 0.08 & - & - & - & - & - \\ 
29521 & 97 & 5129 & 2.46 & -2.29 & 1.6 & -0.11 & 0.10 & 1 & 0.00 & 0.10 & 0.32 & 0.08 & - & - & - & - & - \\ 
41987 & 91 & 5130 & 2.41 & -2.33 & 1.6 & 0.11 & 0.10 & 1 & - & - & 0.30 & 0.08 & - & - & - & - & - \\ 
1939 & 88 & 5149 & 2.48 & -2.41 & 1.6 & 0.02 & 0.10 & 1 & 0.02 & 0.10 & -0.06 & 0.10 & - & - & - & - & - \\ 
34272 & 84 & 5151 & 2.47 & -2.25 & 1.6 & 0.10 & 0.10 & 1 & 0.01 & 0.10 & 0.35 & 0.08 & - & - & - & - & - \\ 
35961 & 87 & 5156 & 2.44 & -2.41 & 1.6 & -0.06 & 0.10 & 1 & - & - & -0.41 & 0.10 & - & - & - & - & - \\ 
31010 & 81 & 5164 & 2.46 & -2.31 & 1.6 & 0.04 & 0.10 & 1 & 0.04 & 0.10 & 0.13 & 0.10 & - & - & - & - & - \\ 
23216 & 67 & 5216 & 2.57 & -2.32 & 1.6 & 0.07 & 0.10 & 1 & - & - & -0.48 & 0.10 & - & - & - & - & - \\ 
18240 & 67 & 5229 & 2.59 & -2.30 & 1.6 & 0.32 & 0.10 & 1 & 0.06 & 0.10 & 0.18 & 0.10 & - & - & - & - & - \\ 
32942 & 62 & 5236 & 2.61 & -2.25 & 1.6 & 0.30 & 0.18 & 1 & - & - & 0.38 & 0.10 & - & - & - & - & - \\ 
32979 & 64 & 5236 & 2.57 & -2.35 & 1.6 & 0.00 & 0.13 & 1 & 0.11 & 0.25 & 0.11 & 0.10 & - & - & - & - & - \\ 
2411 & 55 & 5239 & 2.65 & -2.29 & 1.6 & - & - & - & - & - & -0.03 & 0.10 & - & - & - & - & - \\ 
23153 & 70 & 5242 & 2.59 & -2.26 & 1.6 & -0.01 & 0.13 & 1 & - & - & -0.09 & 0.10 & - & - & - & - & - \\ 
8927 & 61 & 5250 & 2.64 & -2.42 & 1.6 & 0.12 & 0.15 & 1 & - & - & -0.34 & 0.10 & - & - & - & - & - \\ 
18770 & 63 & 5260 & 2.63 & -2.38 & 1.6 & 0.19 & 0.15 & 1 & - & - & 0.03 & 0.10 & - & - & - & - & - \\ 
22441 & 53 & 5270 & 2.70 & -2.35 & 1.6 & 0.15 & 0.15 & 1 & - & - & 0.23 & 0.13 & - & - & - & - & - \\ 
18508 & 46 & 5276 & 2.68 & -2.26 & 1.6 & 0.35 & 0.15 & 1 & - & - & -0.01 & 0.13 & - & - & - & - & - \\ 
9608 & 50 & 5291 & 2.68 & -2.33 & 1.6 & 0.13 & 0.15 & 1 & - & - & -0.29 & 0.10 & - & - & - & - & - \\ 
29436 & 57 & 5312 & 2.81 & -2.41 & 1.5 & 0.26 & 0.15 & 1 & - & - & 0.04 & 0.10 & - & - & - & - & - \\ 
4989 & 58 & 5319 & 2.72 & -2.38 & 1.6 & 0.06 & 0.13 & 1 & - & - & 0.18 & 0.10 & - & - & - & - & - \\ 
3825 & 50 & 5323 & 2.76 & -2.26 & 1.5 & 0.11 & 0.13 & 1 & - & - & 0.52 & 0.10 & - & - & - & - & - \\ 
40762 & 58 & 5325 & 2.78 & -2.34 & 1.5 & 0.19 & 0.13 & 1 & - & - & 0.32 & 0.13 & - & - & - & - & - \\ 
8329 & 53 & 5344 & 2.86 & -2.41 & 1.5 & 0.41 & 0.18 & 1 & - & - & 0.09 & 0.13 & - & - & - & - & - \\ 
2463 & 59 & 5349 & 2.84 & -2.31 & 1.5 & 0.16 & 0.13 & 1 & - & - & 0.39 & 0.10 & - & - & - & - & - \\ 
19346 & 56 & 5349 & 2.80 & -2.34 & 1.5 & 0.19 & 0.15 & 1 & - & - & 0.62 & 0.13 & - & - & - & - & - \\ 
28026 & 47 & 5362 & 2.81 & -2.38 & 1.5 & 0.14 & 0.18 & 1 & - & - & -0.20 & 0.13 & - & - & - & - & - \\ 
31796 & 40 & 5374 & 2.87 & -2.25 & 1.5 & 0.20 & 0.20 & 1 & - & - & -0.17 & 0.13 & - & - & - & - & - \\ 
\\
\hline \\
\end{longtable}
\renewcommand{\arraystretch}{1.00}
\end{center}
\end{landscape}
\twocolumn

\end{document}